\documentclass{aa}  

\usepackage{graphicx}
\usepackage{amsmath} % for \text command
\usepackage{booktabs} % for better horizontal rules
\usepackage{threeparttable} % for table notes
\usepackage{txfonts}
\newcommand{\tablenotemark}[1]{\textsuperscript{#1}}
\usepackage{caption}
\usepackage{longtable}
\usepackage{xtab}
\usepackage{tablefootnote}
\usepackage{multirow}
\usepackage{float}
\usepackage{natbib} % Include the natbib package
\usepackage{subfigure}
\usepackage{subcaption}
\usepackage{afterpage}
\usepackage{url}
\usepackage{hyperref}
\usepackage{placeins} % Provides \FloatBarrier command
\hypersetup{
    colorlinks=true,
    linkcolor=blue,
    citecolor=blue,
    urlcolor=blue
}
\begin{document} 

\title{Spectral dataset of young type Ib supernovae and their time evolution}
\titlerunning{Spectral dataset of young type Ib SNe}

% \author[]{N. Yesmin}
% \affil{Department of Astronomy, University of Virginia, Charlottesville, VA 22904, USA}

% \author[0000-0002-7472-1279]{C. Pellegrino}
% \affil{Department of Astronomy, University of Virginia, Charlottesville, VA 22904, USA}

% \author[0000-0001-7132-0333]{M. Modjaz}
% \affil{Department of Astronomy, University of Virginia, Charlottesville, VA 22904, USA}
   \author{N. Yesmin
          \inst{1}
          \and
           C. Pellegrino\inst{1} 
          \and
         M. Modjaz\inst{1}
          \and
          R. Baer-Way \inst{1}
          \and 
          D.~A. Howell\inst{2,3}
          \and 
          I. Arcavi\inst{4}
          \and
          J. Farah\inst{2,3}
          \and
          D. Hiramatsu\inst{5,6} %ORCiD: 0000-0002-1125-9187
          \and
          G. Hosseinzadeh\inst{7,8}
        \and
          C. McCully\inst{2,3}
          \and 
          M. Newsome\inst{2,3}
          \and
          E. Padilla Gonzalez\inst{2,3}
          \and
          G. Terreran\inst{2,3} 
          \and
          S. Jha\inst{9}
  }
  \authorrunning{N. Yesmin et al.}
    %      \email{pqt7tv@virginia.edu}
    % \and
    %     \email{mmodjaz@virginia.edu}
  \institute{Department of Astronomy, University of Virginia, Charlottesville, VA 22904, USA\\
    \email{pqt7tv@virginia.edu, vru7qe@virginia.edu}
    \and
    Las Cumbres Observatory, 6740 Cortona Drive, Suite 102, Goleta, CA 93117-5575, USA
    \and
    Department of Physics, University of California, Santa Barbara, CA 93106-9530, USA
    \and 
    School of Physics and Astronomy, Tel Aviv University, Tel Aviv 69978, Israel
    \and 
    Center for Astrophysics, Harvard \& Smithsonian, 60 Garden Street, Cambridge, MA 02138-1516, USA
    \and 
    The NSF AI Institute for Artificial Intelligence and Fundamental Interactions, USA
    \and
    Steward Observatory, University of Arizona, 933 North Cherry Avenue, Tucson, AZ 85721-0065, USA
    \and
    Department of Astronomy \& Astrophysics, University of California, San Diego, 9500 Gilman Drive, MC 0424, La Jolla, CA 92093-0424, USA
    \and
    Department of Physics and Astronomy, Rutgers, The State University of New Jersey, 136 Frelinghuysen Road, Piscataway, NJ 08854-8019, USA
}

 \abstract
  {Due to high-cadence automated surveys, we can now detect and classify supernovae (SNe) within a few days after explosion, if not earlier. Early-time spectra of young SNe directly probe the outermost layers of the ejecta, providing insights into the extent of stripping in the progenitor star and the explosion mechanism in the case of core-collapse supernovae. However, many SNe show overlapping observational characteristics at early times, complicating the early-time classification. In this paper, we focus on the study and classification of type Ib supernovae (SNe Ib), which are a subclass of core-collapse SNe that lack strong hydrogen lines but show helium lines in their spectra. Here we present a spectral dataset of eight SNe Ib, chosen to have at least three pre-maximum spectra, which we call early spectra. Our dataset was obtained mainly by the Las Cumbres Observatory (LCO) and it consists of a total of 82 optical photospheric spectra, including 38 early spectra. This dataset increases the number of published SNe Ib with at least three early spectra by $\sim$60\%. For our classification efforts, we used early spectra in addition to spectra taken around maximum light. We also converted our spectra into SN IDentification (SNID) templates and make them available to the community for easier identification of young SNe Ib. Our dataset increases the number of publicly available SNID templates of early spectra of SNe Ib by $\sim$43$\%$. Half of our sample has SN types that change over time or are different from what is listed on the Transient Name Server (TNS).  We discuss the implications of our dataset and our findings for current and upcoming SN surveys and their classification efforts.}
\maketitle
%----------End Abstract---------------

\section{Introduction}\label{sec:intro}
\par Most massive stars end their life cycles as core-collapse supernovae (CCSNe), which are categorized by their spectral features: type I supernovae (SNe) lack H lines, while type II SNe exhibit them. type I CCSNe are further divided into subclasses, which are collectively called stripped envelope supernovae (SESNe, \citealt{1997ApJ...483..675C}; \citealt{1997ARA&A..35..309F,GalYam17};
\citealt{modjaz2019new}), as they have been stripped of their outer layers: type Ib SNe (SNe Ib) are characterized by the presence of strong He lines and the absence of strong H lines (though even SNe Ib may have a weak H$\mathrm{\alpha}$ line \citealt{Liu16}), and type Ic SNe (SNe Ic) are devoid of both H and He lines. Transitional type IIb SNe (SNe IIb) show diminishing H lines over time, but with emerging He lines akin to SNe Ib. Additionally, a subset of high-velocity SNe Ic with broad lines (SNe Ic-bl, e.g., \citealt{2016ApJ...832..108M}) has been interesting because they are the only kind of CCSNe observed to be connected to long-duration gamma-ray bursts (for reviews, see, e.g., \citealt{woosley06,modjaz11,cano17}). %Stripped envelope supernovae (SESNe) constitute a CCSNe subclass, involving massive stars that shed their envelopes before explosion; thus, types Ib, IIb, and Ic fall into the SESNe category (see \citeauthor{modjaz2019new} \citeyear{modjaz2019new} for details). 

%--------- BEGIN TABLE 1 -------------
\begin{table*}[!htb]
\captionsetup{labelfont=bf}
%\captionsetup[table]{labelfont=bf}
\caption{Overview of SNe Ib included in our sample.} 
\label{tab:sn_summary}
\centering
\begin{tabular}{lccccc}
\hline\hline
\textbf{SN name} & \textbf{SN type\tablenotemark{a}} &  \textbf{z}\tablenotemark{b} & \textbf{N$_{spec}$\tablenotemark{c}} & \textbf{\parbox{12em}{\centering Date of maximum light in\\V-band $t_{V_{\text{max}}}$ (MJD)}} & \textbf{\parbox{10em}{\centering Statistical uncertainty\\$\Delta t_{V_{\text{max}}}$ (MJD)}}\\
\hline
SN 2023ljf & Ib & 0.015 &  9 &  60132.87 & 0.31 \\
\midrule
SN 2022nyo & IIb -> Ib\tablenotemark{e} & 0.010637 & 5 & 59776.91 & 0.29 \\
\midrule
SN 2021ukt\tablenotemark{d} & IIn -> Ib\tablenotemark{f}  & 0.012886 & 12 & 59441.66 & 0.30 \\
\midrule
SN 2021njk & Ib &  0.012952 & 8 &  59375.75 & 0.18 \\
\midrule
SN 2021hen\tablenotemark{d} & Ib & 0.0211  & 6 & 59311.73 & 0.27 \\
\midrule
SN 2020hvp & Ib & 0.005247 & 16 & 58975.95 & 0.46\\
\midrule
SN 2019odp & Ic-bl -> Ib\tablenotemark{g} & 0.014353 & 13 & 58735.19 & 0.16\\
\midrule
SN 2016bau & Ib & 0.003856 & 13 & 57478.17 & 0.69\\
%SN 2015ap & Ib & 0.011 & 57285.47 & 0.71 \\
\hline
\end{tabular}
\tablefoot{
\tablenotemark{a} Each SN is classified by running the spectrum closest to the V-band maximum and the available earliest spectrum through SNID (see Section \ref{sec:classification} for details) with the most updated SESNe template library (\citealt{modjaz2014optical, Liu16, Liu17, Williamson2019, Williamson2023YoungSNIc}).\\
\tablenotemark{b} Redshift values are obtained from TNS, except for SN 2022nyo, for which the redshift was assumed to be the redshift of its host galaxy available on NASA/IPAC Extragalactic Database (NED; \citealt{Helou_1991_NED}).\\
\tablenotemark{c} The total number of photospheric spectra for the corresponding SN included in this work. \\
\tablenotemark{d} For these 2 SNe, the LCO V-band light curves do not capture the rise to the peak. Therefore, the i-band light curve is used to calculate the date of the peak and is later converted to the date of the peak in the V-band using Table 10 of \cite{bianco2014multi}, assuming these SN lightcurves have the same behavior of the SESNe in \cite{bianco2014multi}. Conversion of the date of the peak in the i band to the date of the peak in the V band adds a systematic uncertainty of $\sim$1.5 days. We note that the additional systematic uncertainty of $\sim$1.5 days from converting from i-band to V-band is not listed in the entries above. Additionally, using the conversion relationship between the date of maximum in the V-band and i-band from \cite{Rodríguez_2023_Iron-yield} (see Section \ref{sec:peak_calc}), we obtain $t_{V_\text{max}}$ = 59442.06 $\pm$ 1.63 and 59312.13 $\pm$ 1.62 for SNe 2021ukt and 2021hen respectively.\\
\tablenotemark{e} The earliest spectrum of SN 2022nyo, obtained on UT 2022-07-03 01:55:28 (with $t_{V_\text{max}}$ = $-$13.8 days), shows a prominent hydrogen-like feature around 6200 $\AA$. This characteristic signature of a hydrogen line is further validated by the SNID results, which identify the best match with SN 2008cw (type IIb). However, subsequent spectra acquired on UT 2022-07-12 18:29:14 (with $t_{V_\text{max}}$ = $-$4.1 days) display spectral characteristics consistent with SNe Ib (see Section \ref{subsec: 22nyo} for further details).\\
%A subset of high-velocity SNe Ic with broad lines (SNe Ic-bl) are astrophysically interesting because they are the only kind of SNe observed to be connected to gamma-ray bursts (GRBs) \citep{2016ApJ...832..108M}. \\
\tablenotemark{f} SN~2021ukt is a peculiar SN Ib with IIn-like features (see the section \ref{subsec: 21ukt} about this SN): at early times, the TNS spectrum showed only H$\alpha$ and no He absorption lines, thus the SN IIn classification above, but over times the multiple He absorption lines arise clearly, thus the SN Ib classification, while the narrow H$\alpha$ line persisted at weaker, but varying, levels of strength. \\
\tablenotemark{g} SNID initially classifies SN 2019odp as broad-line type Ic (Ic-bl) based on its spectra taken before the V-band maximum light ($t_{V_\text{max}}$ < 0), but later classifies it as type Ib based on the spectra taken after the V-band maximum light ($t_{V_\text{max}}$ > 0). See Section \ref{subsec: 19odp} for more information.\\ %2019odp was initially classified as type Ic-bl by \cite{2019TNSCR1595....1B} on August 23, 2019 (MJD = 58718.2) during the ePESSTO+ survey \citep{2015A&A...579A..40S}. According to \cite{Schweyer_2023_19odp}, the early spectrum of SN 2019odp, taken 16 days before the g-band peak ($t_{g}_{max}$ = -16d), matched type Ic-bl. Later, they reclassified it as type Ib based on the spectrum obtained 26 days after the g-band peak ($t_{g}_{max}$ = 26d). Thus, our pre-maximum and post-maximum classification of SN 2019odp confirms the classification by \cite{Schweyer_2023_19odp}.%
}
\end{table*}
%--------------END TABLE 1------------------
\setlength{\parskip}{0pt}

\par Despite the understanding that SESNe result from the core collapse of massive stars, the diverse range of observed explosion properties poses a challenge in classifying SESNe into distinct types. As a result, it remains unclear if the classification types are distinct or share overlapping characteristics, and what properties of the progenitors or of the explosion cause this diversity in the observed SNe properties. Addressing these unresolved questions involves two approaches: (i) analyzing data of large samples to identify trends in different SN types and (ii) studying data of SNe well before maximum light to directly probe the outermost layers of the progenitor envelope, as the photosphere lies in the outermost layers of the ejecta during this early phase and recedes into deeper layers over time. 
\par Traditionally, individual SNe have been thoroughly studied, such as type IIb SN 1993J (\citealt{filippenko1993type,matheson2000optical,matheson2000detailed}) and type Ic SN 1994I (\citealt{filippenko1995type,richmond1996ubvri}). The prevalence of population studies has increased in recent years. Statistical comparisons of SESNe spectra by \cite{matheson_2001_sample} revealed heterogeneous properties, while subsequent studies by \cite{Fremling_2018_SESNe}, \cite{Prentice_2019_SESNe}, \cite{Liu16}, and \cite{Holmbo_2023_CSP} explored the spectral properties of different types of SESNe. To quantify and visualize the continuum of SESNe classification, \cite{Williamson2019} presented a new quantitative and data-driven classification technique utilizing the machine learning tool of the support vector machine (SVM) after applying principal component analysis (PCA) to the photospheric spectra of SESNe. They find that the SESNe types, in particular SNe Ib, are most distinguishable in later phase ranges, particularly 10 $\pm$ 5 days and 15 $\pm$ 5 days relative to the V-band maximum. %However, the dataset included in \cite{2019ApJ...880L..22W} lacked many pre-maximum spectra for SNe Ib, as it included data from earlier surveys, which had low cadences. 

\par However, these studies lack early spectra in their sample because they include data from earlier surveys that have low cadences. Advancements in high-cadence wide-field surveys now enable prompt detections and observations of SNe to be made. Through these early time observations, though, we are discovering an increasing number of SNe with evolving properties, such as SNe that change their spectroscopic type throughout their evolution. These SNe pose a challenge to our classification efforts. The transitional SN 2022crv \citep{Dong_2024_22crv,Gangopadhyay_2023_22crv} is an example of an evolving SESN, transitioning from type IIb to type Ib in the early time.

\par Now is the perfect time to start conducting comprehensive population studies of young SESNe, in particular SNe Ib. This work constitutes the first step in these population studies, as it publishes optical spectra and spectral templates of young SNe Ib only days after explosion, well before maximum light. Studying the spectra of young SNe Ib over time will provide valuable insights into when the tell-tale He lines will appear. Consequently, it will offer significant constraints on $^{56}$Ni mixing, as the He lines result from nonthermal excitation induced by fast electrons that are accelerated by energetic megaelectron-volt photons emitted during Ni decay \citep{Lucy_1991_nonLTE}. Tracking the appearance of the He lines helps with inferring the distribution of Ni-mixing further out in the ejecta. In turn, the exact details of the Ni mixing can constrain explosion models, which, for objects where the He lines are visible shortly after explosion, would require the release of Ni close to the outermost He layer, potentially indicating large-scale turbulence (e.g., \citealt{hammer10_mixing}).  Furthermore, analyzing early spectra will inform the most appropriate timing to schedule spectroscopy to classify an SN Ib, which is essential for the new era of big-data transient science, in which thousands of SNe will be photometrically discovered every night with the Vera C. Rubin Observatory as part of the Legacy Survey of Space and Time (LSST; \citealt{Ivezi_2019_LSST}). 
% a layout of what is presented in this paper
\par In Section \ref{sec:observations}, we present a dataset of 82 photospheric spectra of eight different SNe Ib, including 38 spectra taken before maximum light. All spectra will be made public via the Weizmann Interactive Supernova Data REPository (WISeREP\footnote{\url{https://www.wiserep.org/}}; \citealt{Yaron_2012_wiserep}). We discuss our spectrum phase calculation and the method for calculating date of maximum light in Section \ref{sec:peak_calc}. In Section \ref{sec:classification}, we discuss the specifics of the SuperNova IDentification (SNID) code \citep{Blondin&Tonry_2007_SNID} classification of each SN in the sample. In Section \ref{sec:snid_templates}, we present our procedure for creating spectral templates for SNID for the SNe in our sample using our data as well as prior published data. Our SNID templates can be downloaded via our METAL GitHub repository\footnote{\url{https://github.com/metal-sn/SESNtemple/tree/master}}, formerly known as the SNYU github page.

%https://github.com/metal-sn/
%--------------------------------------------------------------------
%----------Begin Illustrative Time-series Figure----------------
\begin{figure*}[!htb]
    \centering
    \begin{minipage}[b]{0.45\linewidth}
        \centering
        \includegraphics[width=\linewidth]{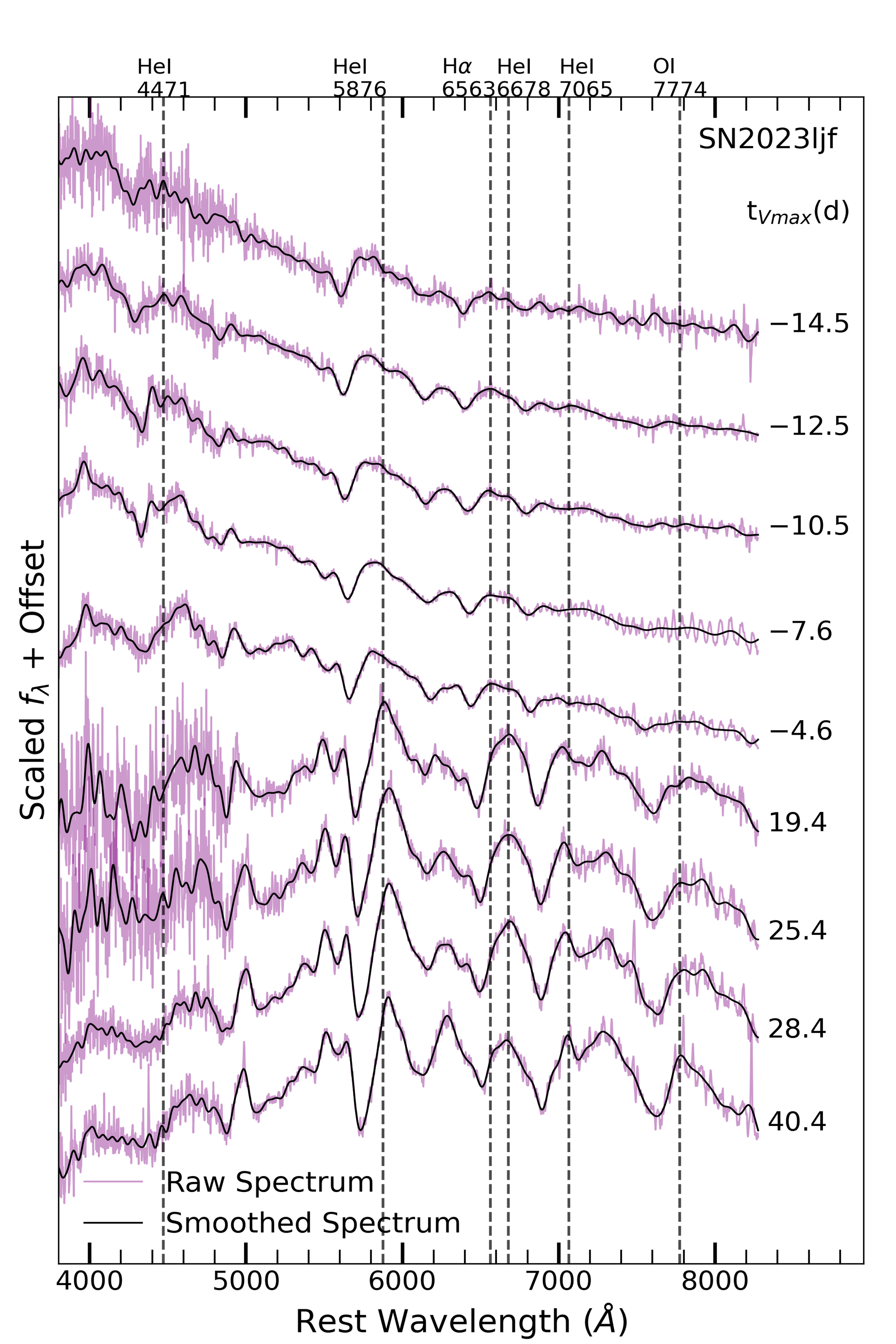}
    \end{minipage}
    \hspace{0.05\linewidth} % Adjust the horizontal space as needed
    \begin{minipage}[b]{0.45\linewidth}
        \centering
        \includegraphics[width=\linewidth]{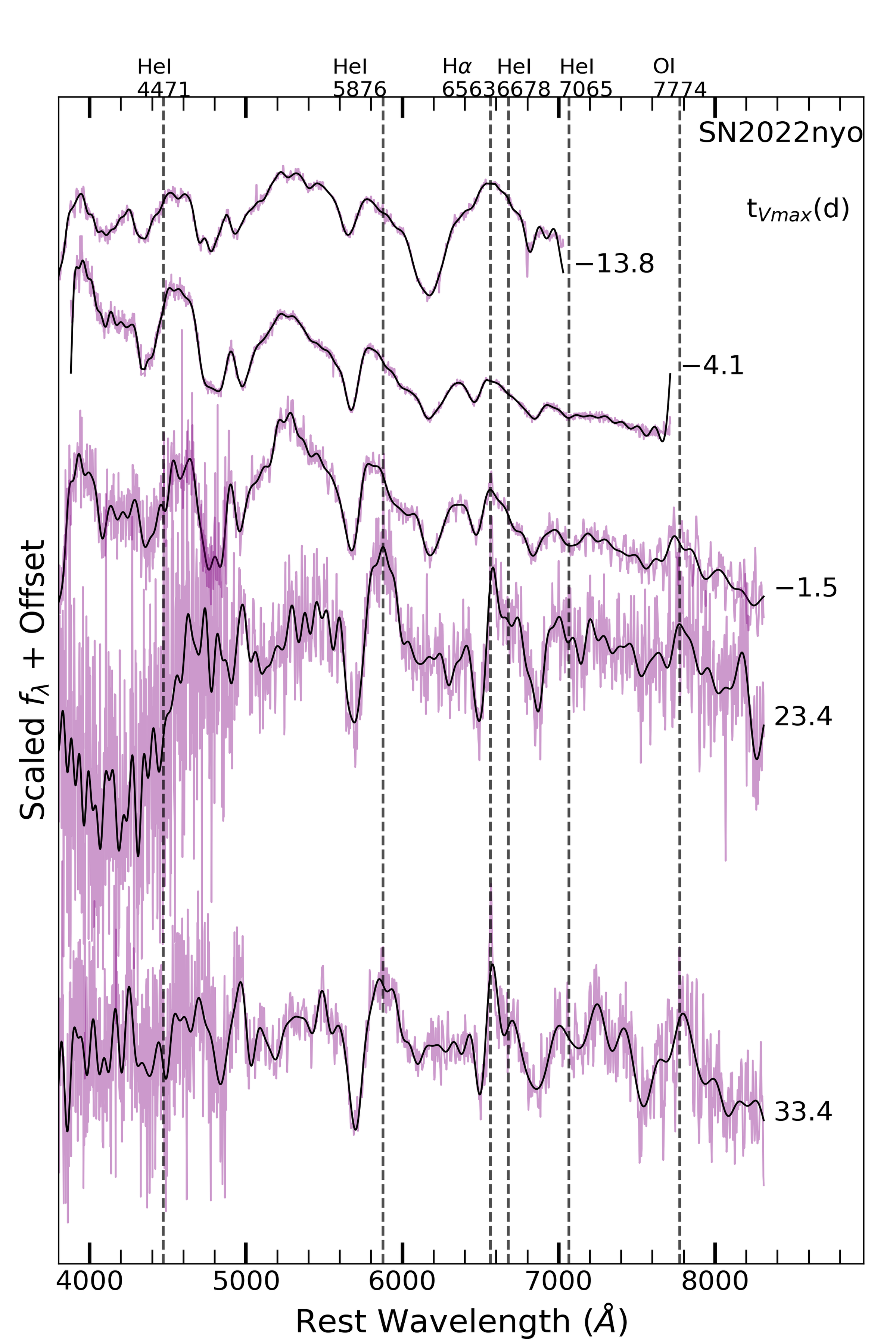}
    \end{minipage}
    \caption{Illustrative spectral time series for two SNe Ib (SN2023ljf and SN2022nyo) presented in this work. The flux \( f_{\lambda} \), measured in units of \(\text{erg} \, \text{s}^{-1} \, \text{cm}^{-2} \, \text{Å}^{-1}\), is normalized, and an offset is added to enhance the visualization of the temporal evolution of the spectra. The smoothed spectra (produced following the procedure described in Appendix B of \citealt{Liu16}), shown as solid black lines, are highlighted in the foreground, while the reduced spectra before smoothing are in purple in the background. Dashed vertical lines mark the HeI 4471, HeI 5876, HeI 6678, HeI 7065, H$\alpha$ 6563, and OI 7774 lines, and the phases are relative to the date of the V-band maximum ($t_{V_\text{max}}$). We note that we manually removed galaxy emission lines from the spectra at $t_{V_\text{max}}$ = $-$13.8d and $-$4.1d of SN 2022nyo. Spectral time-series for all eight SNe Ib is shown in Figure ~\ref{fig:full_spectra}.}
    \label{fig:illustrative_spectra}
\end{figure*}
%\FloatBarrier
% ------------------END Illustrative Time-series Figure---------

% ---- End of Introduction------

\section{Spectroscopic observations and data reduction}\label{sec:observations}
Our sample consists of eight SNe Ib with data taken between 2016 and 2023 mainly by the Las Cumbres Observatory (LCO) \citep{Brown_2013_LCO} as part of the Global Supernova Project (GSP) \citep{Howell_2019_GSP, Howell_2024_GSP}. These eight SNe Ib were selected from the GSP sample because a) they had observed photometry that covers the lightcurve peak in at least one band, so spectral phases could be calculated relative to the date of maximum (see Section \ref{sec:peak_calc} for details), and b) they had unpublished spectra, with at least three spectra before the date of V-band maximum light (t$_{V_\text{max}}$), such that we could study their early-time behavior. Additional data were taken as part of the LCO Supernova Key Project \citep{Howell_2017_SNKeyProject} for SN 2016bau and by GSP members with other observatories. 
\par Thus, we present a total of 82 photospheric (t$_{V_\text{max}}$ < 60 days) optical spectra of eight SNe Ib. We exclude nebular-phase spectra for this study since we focus on photospheric-phase classification. Our dataset includes a total of 38 spectra taken before the V-band maximum light, which we refer to as early spectra, increasing the number of published SNe Ib with at least three early spectra by $\sim$60$\%$ and the number of publicly available early spectra of SNe Ib that satisfy the same criterion of at least three early spectra by $\sim$37$\%$. Table \ref{tab:sn_summary} shows a summary of our dataset. Figure \ref{fig:illustrative_spectra} shows an illustrative example of the spectral time series for two SNe Ib from our sample, while Figure \ref{fig:full_spectra} presents the complete spectral time series for all eight SNe Ib in our study.

\par Almost all (80 out of 82) optical spectra were taken using the FLOYDS spectrographs on the 2m telescopes at Siding Spring Observatory (COJ 2m), Australia, and Haleakala (OGG 2m), Hawaii. A 2\arcsec slit was
placed on the SN along the parallactic angle \citep{Filippenko1982}. The spectra are extracted from the raw images with wavelength scale and flux calibration applied by the FLOYDS pipeline \footnote{\url{https://lco.global/documentation/data/floyds-pipeline/}}, following standard procedures \citep{Valenti14}. One optical spectrum was taken with the Southern African Large Telescope (SALT) using the Robert Stobie Spectrograph
(RSS; \citealt{Smith06}) through Rutgers University program 
2022-1-MLT-004 (PI: Jha). The observations were taken with the PG0900 grating and 1.5\arcsec wide longslit, giving a spectral resolution $R = \lambda/\Delta \lambda \simeq 1000$. The data were reduced using a custom pipeline based on standard Pyraf spectral reduction routines and the PySALT package \citep{Crawford10}. Additionally, one spectrum was obtained with the Goodman High Throughput Spectrograph on the Southern Astrophysical Research (SOAR) 4.1m telescope. The spectrum was taken using the blue camera, covering the wavelength range 3500 -- 7000 \AA{}, with a 1\arcsec slit and 400 mm$^{-1}$ line grating. We used the Goodman Spectroscopic Data Reduction Pipeline to perform bias and flat corrections, cosmic-ray removal, and wavelength calibration. Fluxes were calibrated to a standard star observed on the same night with the same instrumental setup. The details of the observation log are presented in Appendix \ref{sec: obs} Table \ref{tab: obs_log}. All the spectra presented in this work will be made available through WISeREP. 

\par For the identification of the supernova type, we initially relied on the classification available from the Transient Name Server (TNS\footnote{\url{https://www.wis-tns.org/}}; \citealt{Gal-Tam_2021_TNS}), the official mechanism of the International Astronomical Union (IAU) for reporting new astronomical transients such as SN candidates. To further solidify the classification of the SNe, we ran the spectra taken closest to the date of the V-band maximum light and the early spectra through SNID. Three out of eight SNe have classifications that are different from what is listed on the TNS. Details on the classification methodology are discussed in Section \ref{sec:classification}.
%--------------------------------------------------------------------

%-----------BEGIN TABLE 2------------
\begin{table*}[!htb]
\captionsetup{labelfont=bf}
\caption{Comparison of our calculated maximum dates and uncertainties with published values.} 
\label{tab: max_date_comparison}
\centering
\begin{tabular}{lcccccc}
\hline\hline
\textbf{SN name} & \textbf{\parbox{6em}{\centering Our calculated $t_{V_{\text{max}}}$ (MJD)}} & \textbf{\parbox{6em}{\centering Our calculated $\Delta t_{V_{\text{max}}}$ (MJD)}} & \textbf{\parbox{6em}{\centering $t_{V_{\text{max}}}$ from published work (MJD)}} & \textbf{\parbox{6em}{\centering $\Delta t_{V_{\text{max}}}$ from published work (MJD)}} & \textbf{\parbox{4em}{\centering Reference}} & \textbf{\parbox{6em}{\centering Method of conversion \tablenotemark{a}}} \\
\hline
SN 2020hvp & 58975.95 & 0.46 & 58975.7 & 0.4 & \cite{Rodríguez_2023_Iron-yield} & -\\
\midrule
SN 2019odp & 58735.19 & 0.16 & 58735.2\tablenotemark{b} &  1.1\tablenotemark{b} & \cite{Schweyer_2024_19odpClassification} & R1\\
 & & & 58736.1 &  0.6 & \cite{Rodríguez_2023_Iron-yield} & -\\
\midrule
SN 2016bau & 57478.17 & 0.69 & 57479.37\tablenotemark{c} & 2.23\tablenotemark{c} & \cite{Aryan_2021_16bau} & R1\\
& & & 57479.67\tablenotemark{c} & 1.99\tablenotemark{c, d} & \cite{Aryan_2021_16bau} & B10\\
 & & & 57477.7 & 0.2 & \cite{Rodríguez_2023_Iron-yield} & -\\
\hline
\end{tabular}
\tablefoot{
    \tablenotemark{a} R1 = Table 1 from \cite{Rodríguez_2023_Iron-yield}, B10 = Table 10 from \cite{bianco2014multi}, and "-" means no conversion is needed. See Section \ref{sec:peak_calc} for details. \\
    \tablenotemark{b} Converted from $t_{g_{\text{max}}}$. \\
    \tablenotemark{c} Converted from $t_{B_{\text{max}}}$. \\
    \tablenotemark{d} An additional systematic uncertainty of $\sim$1.3 days, resulting from the conversion between B-band and V-band, is not included in the reported uncertainty above. \\
}
\end{table*}
%---------- END TABLE 2---------------

% -------------End of Spectroscopic observations-----------

\section{Spectrum phases and V-band maximum date calculation}\label{sec:peak_calc}
Determining the phase of each spectrum, or the epoch relative to maximum light, is essential for conducting a systematic analysis of different SN properties. Our phase calculations are based on the V-band maximum date ($t_{V_{\text{max}}}$), ensuring consistent tracking of the appearance of He lines across our sample. This approach also facilitates consistent comparisons with other population studies, such as those conducted by \cite{Liu16} and \cite{Holmbo_2023_CSP}, which similarly adopt the time of the V-band maximum date as their reference point for phase calculation.

\par While all SNe in our sample have V-band light curves taken by LCO, 2 SNe lack a clear peak in the V-band light curve. For these two cases, namely SNe 2021hen and 2021ukt, we utilized the LCO i-band light curve to determine the maximum date in the i-band ($t_{i_\text{max}}$), as redder bands exhibit maxima at later times. We then convert the determined maximum date in the i-band to the maximum date in the V-band using Table 10 from \cite{bianco2014multi}, assuming these SN lightcurves have the same behavior as the SESNe in \cite{bianco2014multi}, which for SN~2021ukt is a large caveat (see section~\ref{subsec: 21ukt}). This step adds a systematic uncertainty of $\sim$1.5 days. Alternatively, we could use the average values between $t_{V_{\text{max}}}$
and the epochs of maximum light in other optical bands for SNe Ib, as outlined in Table 1 of \cite{Rodríguez_2023_Iron-yield}. This approach yields a conversion equation for the date of maximum light between the V-band and i-band, given by $t_{V_{\text{max}}}$ = ($t_{i_{\text{max}}}$ + 3.1) $\pm$ 1.6. Using this conversion relationship, the maximum dates for SNe 2021hen and 2021ukt are provided in the footnote of Table \ref{tab:sn_summary}. However, we opt to use the converted dates from \cite{bianco2014multi} for the phase calculations of our spectra to be consistent. See below for more details. 

\par We perform the maximum light date calculations using the light curves after their preliminary reductions; as such, we do not publish the full photometry in this paper. To measure the date of maximum light for the SNe in our sample, we follow \cite{bianco2014multi} and use a Monte Carlo (MC) simulation process to fit a quadratic around the peak of the light curve for a given filter. The adopted date of maximum light is the mean of the maximum epochs across the MC realizations, and the error associated with the date of maximum light is the corresponding standard deviation. Table \ref{tab:sn_summary} shows the dates of maximum in the V band and their associated uncertainties that we calculated for each of the SNe in our dataset. 

\par We compare our calculated maximum dates for several SNe in our sample with previously published results and find agreement in all cases. Table \ref{tab: max_date_comparison} presents a comparison between our calculated maximum dates and those from previously published results. For SN 2019odp, we find a V-band maximum date of Modified Julian Date (MJD) 58735.19 $\pm$ 0.16. In comparison, \cite{Schweyer_2023_19odp} reported a maximum date of MJD 58734 $\pm$ 1 day in the g-band ($t_{g_\text{max}}$), calculated with their own data. To convert the date of maximum in the g-band to the V-band, we use the average value between $t_{V_{\text{max}}}$ and $t_{g_{\text{max}}}$ from Table 1 of \cite{Rodríguez_2023_Iron-yield}, which gives the relationship specifically for SNe Ib: $t_{V_{\text{max}}}$ = ($t_{g_{\text{max}}}$ + 1.2) $\pm$ 0.5. This conversion yields $t_{V_{\text{max}}}$ = 58735.2 $\pm$ 1.1 for SN 2019odp, consistent with our result. Furthermore, \cite{Rodríguez_2023_Iron-yield} determined the V-band maximum for SN 2019odp as MJD 58736.1 $\pm$ 0.6, based on data from both the literature and the Zwicky Transient Facility (ZTF; \citealt{Graham_2019_ZTF}), further supporting the consistency of our measurements. For SN 2020hvp, our determined V-band maximum of MJD 58975.95 $\pm$ 0.46 is in agreement with \cite{Rodríguez_2023_Iron-yield}, who report MJD 58975.7 $\pm$ 0.4, indicating consistency within the uncertainties. For SN 2016bau, \cite{Aryan_2021_16bau} found that the maximum date in the B band ($t_{B_\text{max}}$) was MJD 57477.37 $\pm$ 1.99 days, calculated from their own data. For comparison, we converted the $t_{B_\text{max}}$ to our calculated $t_{V_\text{max}}$ of MJD 57478.17 $\pm$ 0.69 by using Table 10 from \cite{bianco2014multi}, resulting in converted  $t_{V_\text{max}}$ MJD 57479.67 $\pm$ 1.99 days, with an additional systematic uncertainty of $\sim$1.3 days. One caveat here is that this conversion relationship from \cite{bianco2014multi} is calculated for different SESNe subtypes, and we assume that the light curves of the SNe Ib in our sample behave similarly to those in their sample. But to further verify the conversion, we refer to the relationship reported in Table 1 of \cite{Rodríguez_2023_Iron-yield}, as this conversion is specifically calculated for SNe Ib. According to \cite{Rodríguez_2023_Iron-yield}, the conversion relationship is given by $t_{V_{\text{max}}}$ = ($t_{B_{\text{max}}}$ + 2.0) $\pm$ 1.0, which yields $t_{V_{\text{max}}}$ = 57479.37 $\pm$ 2.23, consistent with our result. Our result for SN 2016bau is also in agreement with \cite{Rodríguez_2023_Iron-yield}, who report a V-band maximum of MJD 57477.7 $\pm$ 0.2.

%----Spectrum phases and V-band maximum date calculation--------

\section{Supernova classification via SNID}\label{sec:classification}
\setlength{\parskip}{0pt}

Ensuring a robust classification of SNe in our sample as type Ib is an essential criterion for our dataset. As discussed below in detail for each SN in our sample, we classify each SN by running the spectrum closest to the V-band peak ($t_{V_{\text{max}}}$ = 0 days) through SNID (\citealt{Blondin&Tonry_2007_SNID}). We also run SNID on the early spectra (i.e., $t_{V_{\text{max}}}$ < 0), including the earliest spectrum taken after discovery, to verify consistency and identify SNe showing uncertain or evolving early types. SNID operates by removing the continuum from the newly obtained spectrum (in order to focus on the spectral lines that define the most common SN types and to take out the effect of any reddening) and by cross-correlating it with a library of previously observed and classified SN spectra, called  individual "SN templates". SNID provides a quality parameter called rlap to measure the correlation; generally, correlations with rlap > rlapmin = 5.0 are considered a good match (\citealt{Blondin&Tonry_2007_SNID}).  Therefore, the SNID matches reflect spectral behavior in only the absorption lines, with stronger absorption lines having higher weights, and do not take into account the shape of the continuum. For the SNID SESN template libraries, we use the most up-to-date ones based on the Modjaz Group Sample (MGS; \citealt{modjaz2014optical, Liu16, Liu17, Williamson2019, Williamson2023YoungSNIc}) to ensure a robust classification of SNe in our sample.

\par Classifying early spectra has been difficult due to the limited availability of previously observed young SNe Ib templates in the SNID library, or in the libraries of  any of the other SN identification codes, such as Superfit \citep{Howell05_Superfit} and GELATO \citep{Gelatoref08}, for that matter. For SNID, the current MGS so far includes 235 spectral templates of 22 SNe Ib with $t_{V_{\text{max}}}$ < 60 days, but only 91 spectra have $t_{V_{\text{max}}}$ < 0 days. Additionally, spectra may not develop strong identifiable features, which is a well-known case for the He lines (e.g., \citealt{Liu16,Williamson2019,williamson2021modeling}). Therefore, while the spectra closest to the V-band maximum light of the SNe in our sample all match those of normal SNe Ib in the SNID library (since they were chosen as such to be included in this paper), some of them may have uncertain or evolving-types before maximum light, which we are uncovering in detail in this paper. For example, as discussed below for the individual SNe, the earliest spectra of SN 2023ljf, SN 2021ukt, and SN 2021njk do not strongly match any specific SN type in SNID. Additionally, SN 2022nyo, SN 2021ukt, and SN 2019odp show evolving types in their early spectra, as described in detail below. Furthermore, we compared our classification with the TNS classification attempts: Table \ref{tab:tns} provides the TNS classification in detail for each SN in the sample. SN 2021ukt and SN 2021njk have different classifications than type Ib on TNS. Below, we detail the discovery and classification information for each of the SNe in our sample. For our classification information, we report the SNID classification results for the earliest spectrum and the spectrum closest to the V-band maximum light, unless otherwise mentioned. 

\par The types of all the spectra in our sample are listed in Table \ref{tab: obs_log}. However, we note that the types for the spectra in Table \ref{tab: obs_log} are not always determinable by SNID due to the limitations of the current SNID library, particularly its incompleteness regarding early spectra (as discussed above). To enhance the reliability of our classifications, we employ complementary methods, including visual confirmation of key spectral features and the fitting of line profiles, as demonstrated in the case of SN 2021ukt (Section \ref{subsec: 21ukt}). We hope that our newly added SNID templates (see Section \ref{sec:snid_templates}) will help address some of the incompleteness in the library regarding early spectra of SNe Ib.

% Description on each SN
\subsection{SN 2023ljf}
SN 2023ljf was discovered by the Asteroid Terrestrial Impact Last Alert System (ATLAS; \citealt{2018PASP..130f4505T}) on UT 2023-06-22 07:40:30 (MJD 60117.32) with a non-detection two days prior \citep{Tonry2023_ljf}. \cite{Pellegrino_2023_23ljfClassification}, as part of the GSP, obtained a classification spectrum of the object on UT 2023-06-23 09:29:05 (t$_{V_\text{max}}$ = -14.5d) and used it to classify SN 2023ljf as a young SN Ib on TNS. We ran our spectrum closest to the V-band maximum light taken on UT 2023-07-03 07:23:36 (t$_{V_\text{max}}$ = $-$4.6d) through SNID. SNID found that the best type is a normal SN Ib, with 55.7$\%$ of the templates with rlap $\ge$ rlapmin corresponding to this type. We also ran our earliest spectrum taken on UT 2023-06-23 09:29:05 (t$_{V_\text{max}}$ = $-$14.5d) through SNID. SNID did not find any favored type or subtype for this spectrum. It is important to note that \cite{Pellegrino_2023_23ljfClassification} used the same spectrum to classify SN 2023ljf as a normal SN Ib on TNS. However, they do not mention the use of SNID for classification. For the subsequent spectrum taken on UT 2023-06-25 08:16:46 (t$_{V_\text{max}}$ = $-$12.5d), SNID found the best match to be SN Ib, with 74.6$\%$ of the templates with rlap $\ge$ rlapmin corresponding to normal SN Ib.

\subsection{SN 2022nyo: Transitioning from SN IIb to SN Ib}\label{subsec: 22nyo} 
SN 2022nyo was discovered by the Distance Less Than 40 Mpc (DLT40) survey \citep{Sand_2023_DLT40} on UT 2022-06-30 06:22:01 (MJD 59760.27) with a non-detection 5 days prior \citep{Pearson2022_nyo}. On TNS, 2022nyo is a type Ib based on the classification of \cite{2022TNSCR2031....1P}, who used a spectrum taken on UT 2022-07-01 00:39:45 (MJD 59761.03) with the Gemini South telescope. We ran our spectrum closest to the V-band maximum light, taken on UT 2022-07-15 10:06:09 (t$_{V_\text{max}}$ = $-$1.5d) through SNID. SNID did not find a favored type match for this spectrum but found the best matches with normal SN Ib. However, the earliest spectrum of SN 2022nyo (t$_{V_\text{max}}$ = $-$13.8d) shows a prominent H$\alpha$ absorption feature around 6200 $\AA$ (implying an absorption velocity of $\sim -$ 20,000 km/s, which is similar to other SNe IIb and SNe Ib at similar phases, see Fig. 2 in \citealt{Liu16}), which diminished in strength in the subsequent spectra starting at $t_{V_\text{max}}$= $-$4.1 days. SNID confirms the H-like feature in the earliest spectrum by finding its best match with the type IIb SN 2008cw \citep{modjaz2014optical} and 87.5$\%$ of the templates with rlap $\ge$ rlapmin corresponding to type IIb. Although the feature at 6200 $\AA$ could be due to other lines, such as C II 6580, Ne I 6402, or Si II 6355 (e.g., \citealt{branch2002direct, 2005ApJ...633L..97T,elmhamdi2006hydrogen,2007PASP..119..135P}), several studies suggest that there is most likely some H in SNe Ib (e.g., \citealt{Liu16}). A qualitatively similar early-time transition from type IIb to type Ib has been noted in SN 2022crv \citep{Dong_2024_22crv, Gangopadhyay_2023_22crv}, though a similarly detailed study of SN~2022nyo is outside the scope of this paper. 
%---------------END SN 2022nyo--------------------

%--------Begin Table 3-------------
\begin{table*}[!htb]
\captionsetup{labelfont=bf}
\caption{SN type at initial classification (e.g., on TNS) and if different, our classification.} 
\label{tab:tns}
\centering
\setlength{\tabcolsep}{4pt}
\begin{tabular}{lccccc}
\hline\hline
\textbf{SN name} & \textbf{\parbox{6em}{\centering Our\\classification}} & \textbf{\parbox{6em}{\centering TNS\\classification}} & \textbf{\parbox{10em}{\centering MJD of TNS\\classification spectrum}} & \textbf{\parbox{8em}{\centering Phase\tablenotemark{a} (days) of TNS\\classification spectrum}} & \textbf{\textbf{Reference for initial classification}}\\
\hline
SN 2023ljf & Ib & Ib & 60118.40 & -14.5 & \cite{Pellegrino_2023_23ljfClassification}\\
\midrule
SN 2022nyo & IIb -> Ib\tablenotemark{b}  & Ib & 59761.03 & -15.9 & \cite{2022TNSCR2031....1P}\\
\midrule
SN 2021ukt & IIn -> Ib\tablenotemark{c} & IIn & 59427.58 & -14.1 & \cite{2021TNSCR2645....1H}\\
\midrule
SN 2021njk & Ib & Ib-pec & 59364.41 & -11.3 & \cite{2021TNSCR1876....1H}\\
\midrule
SN 2021hen & Ib & I & 59303.54 & -8.2  & \cite{Pellegrino2021_SN2021hen_Classification1}\\
& & Ib & 59306.00 & $-$5.7 & \cite{2022TNSCR1989....1S}\\
\midrule
SN 2020hvp & Ib & Ib & 58962.74 & -13.2 & \cite{2020TNSCR1129....1B}\\
& & Ib & 58964.35 & -11.6 & \cite{2020TNSCR1523....1D}\\
\midrule
SN 2019odp & Ic-bl -> Ib\tablenotemark{d} & Ic-bl & 58718.22 & -17.0 & \cite{2019TNSCR1595....1B}\\
& & Ic-bl & 58719.89 & -15.3 & \cite{2019TNSCR2847....1F}\\
& & Ib & 58759.93 & 24.7 & \cite{Schweyer_2024_19odpClassification}\\
% ours on 2019-09-10 13:47:02 which is around MJD 58736
\midrule
SN 2016bau & Ib & Ib & 57461.97 & -16.2 & \cite{2016TNSCR.224....1B}\\
\hline
\end{tabular}
\tablefoot{
    \tablenotemark{a} Phases are calculated with reference to the V-band maximum light. \\
    \tablenotemark{b} Based on our spectra, SN~2022nyo transitioned from a SN IIb to a SN Ib between $t_{V_{\text{max}}}$ = $-$13.8 and $-$4.1 days (see Table~\ref{tab: obs_log}). \\
    \tablenotemark{c} Based on our spectra, SNID identified SN~2021ukt as a SN IIn at $t_{V_{\text{max}}}$ = $-$12.2 days, but as a SN Ib for the subsequent spectrum at $t_{V_{\text{max}}}$ = $-$9.1 days (see Table~\ref{tab: obs_log}). Nevertheless, we note that for many of its spectra, SN~2021ukt seems to have simultaneous SNe Ib- and SN IIn-like features, which has never been seen before—see Section~\ref{subsec: 21ukt}. \\
    \tablenotemark{d} Based on our spectra, SN~2019odp transitioned between a SN Ic-bl and a SN Ib between $t_{V_{\text{max}}}$ = $-$2.7 and 1.4 days (see Table~\ref{tab: obs_log}). \\
}
\end{table*}

%-----------End Table 3-----------------------------
%SN 2021hen which is classified as type I on TNS. Using Superfit (Howell et al. 2005, ApJ, 634, 1190) and SNID, \cite{2021TNSCR1876....1H} found best fits to several type I SNe, including type Ia SN 1999ee and type Ib SN 1998dt, around maximum. It is possible that they did not use MGS. We have the classification spectrum in our sample since it was taken by LCO. We ran the spectrum through SNID and found best match with SN 2004gq, a normal type Ib \cite{modjaz2014optical}. 

\subsection{SN 2021ukt: Peculiar SN Ib with IIn-like features} \label{subsec: 21ukt}
SN 2021ukt was discovered via the ZTF on UT 2021-07-31 10:34:47 (MJD 59426.44) with a non-detection 7 days prior \citep{De2021_ukt}. SN 2021ukt is a unique case within our sample given its peculiar evolution. TNS lists SN 2021ukt as a type IIn due to the classification report of \citet{2021ukt_class} as part of the Spectroscopic Classification of Astronomical Transients (SCAT; \citealt{21ukt_SCAT}) survey, which was based on a quick-reduction spectrum at t$_{V_\text{max}}$=$-$14.1 days (1.9 days before our first spectrum) showing a strong blue continuum and narrow H$\alpha$ and H$\beta$ emission lines. For our earliest spectrum taken on UT 2021-08-03 11:41:04 (t$_{V_\text{max}}$ = $-$12.2d), SNID did not find any significant classification type for this spectrum - nevertheless, a narrow H$\alpha$ that is better fit by a Lorentzian shape ($\chi_{v}^{2}$=1.7) vs. a Gaussian($\chi_{v}^{2}$=2.1) with a FWHM $1640_{-130}^{+140}$ km/s (see below for the fit) is visible, along with its blue continuum, fulfilling the classification criterion of a SN IIn.  For the spectrum on UT 2021-08-06 12:17:32 (t$_{V_\text{max}}$ = $-$9.1d), SNID found matches to 19 SN templates with rlap $\ge$ rlapmin, without a strong match to any single type. Nevertheless, the beginning of the He absorption lines (at $\sim -$15,000 km/s) are seen in the spectra. For the following spectrum taken on UT 2021-08-09 11:57:22 (t$_{V_\text{max}}$ = $-$6.2d), SNID found the favored type to be normal SN Ib, with 60.0$\%$ of the templates with rlap $\ge$ rlapmin corresponding to this type. In addition, we ran our spectrum closest to the V-band maximum, taken on UT 2021-08-15 13:30:07 (t$_{V_\text{max}}$ = $-$0.1d) through SNID. SNID found that the best-fitting type is a normal SN Ib, with 71.1$\%$ of the templates with rlap $\ge$ rlapmin corresponding to normal SN Ib. Thus, we classify SN 2021ukt as a type Ib based on the classification closest to the V-band maximum and mention its type change in Tables~\ref{tab:sn_summary} and \ref{tab:tns}. 

\par
\begin{figure*}
    \centering
    \includegraphics[width=20 cm, height=9 cm]{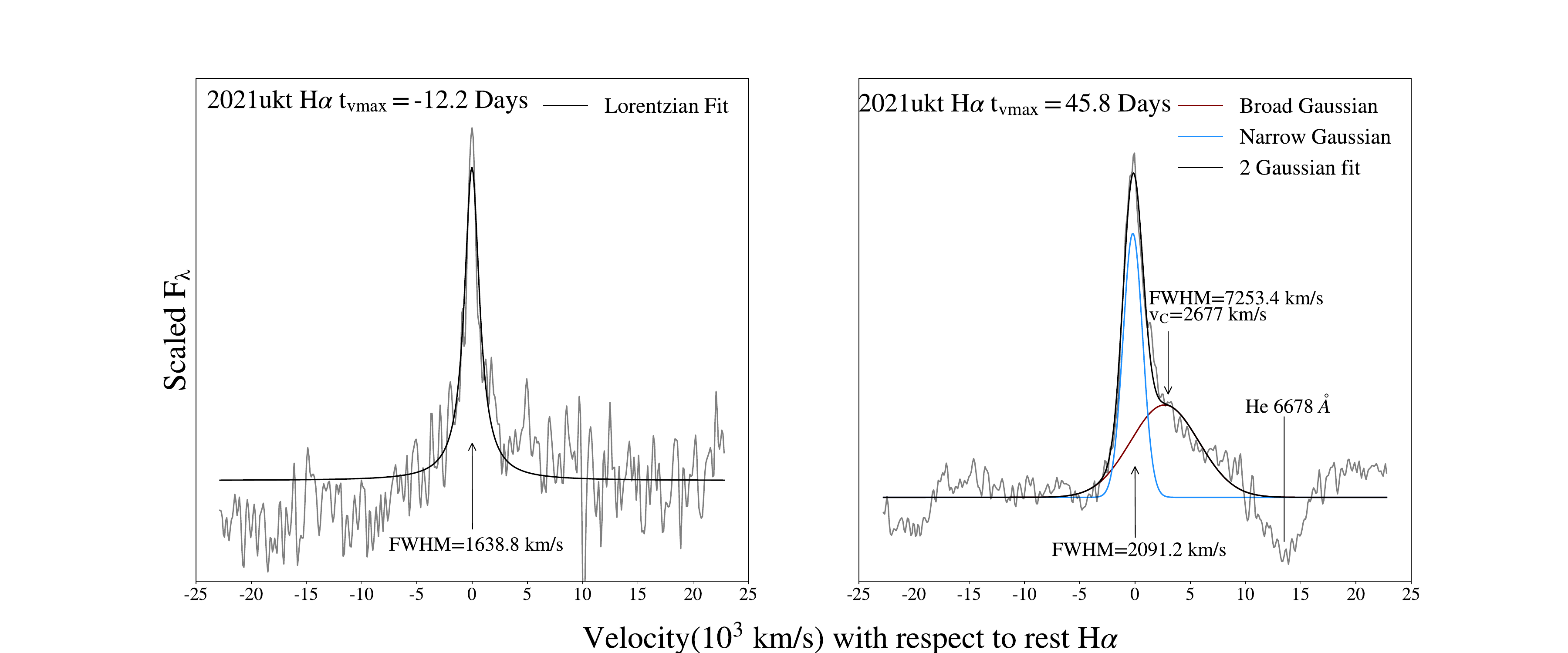}
    \caption{Zoom-in of the $\mathrm{H\alpha}$ emission line of SN~2021ukt at two epochs: the earliest one for which we have a spectrum ($-$12.2 days before $V$-band maximum) and one of the latest. The results of the MCMC fitting performed on continuum-subtracted spectra are shown, with $\mathrm{v_{C}}$ referring to the central velocity of the broad Gaussian in the second spectrum. These careful fits to the $\mathrm{H\alpha}$ emission lines are important for the classification of SN~2021ukt as a SN Ib with IIn features - and at the earliest time, as a bona fide SN IIn.}
    \label{fig:21ukt_ha}
\end{figure*}

 SN 2021ukt displays a broad HeI 5876$\AA$ absorption line which decreases in absorption velocity to $\sim -$8000 km/s at 45.6 days, while the narrow H emission line persists throughout our spectra at varying strengths. As shown in Figure \ref{fig:21ukt_ha}, the shape of the H$\alpha$ emission line evolves from a Lorentzian profile in our earliest spectrum, at $t_{Vmax}$= $-$12.2 days with FWHM $1640_{-130}^{+140}$ km/s to a two-component Gaussian profile in our latest spectrum at $t_{Vmax}$= +45.8 days with a similarly narrow component( FWHM $2090_{-46}^{+47}$ km/s). There is also an intriguing redshifted broad H$\alpha$ component in the spectrum at $t_{Vmax}$= +45.8 days centered at $\sim$ 2680 km/s, but its detailed interpretation is beyond the scope of this work. We obtained fits to the H$\alpha$ emission line by performing MCMC fits (10000 steps with a 1000-step burn-in run) with \texttt{emcee} \citep{Foreman_Mackey_2013} and adopted the 16th and 84th percentiles of the posterior distribution as the upper and lower error bars on the FWHM. 

%---------Begin SN 2021ukt Emission Line Figure-------------

%---------End SN 2021ukt Emission Line Figure-------------

\par The combination of broad He absorption with a narrow H emission line at the same time is something that has never been seen up to this point for an SN of any type. While SN 2014C \citep{Milisavljvevic_2014} has been known to evolve from a SN Ib to an SN IIn and develop strong narrow emission lines, there has never been a case when a SN so strongly displayed characteristics of both subtypes at the same time. A detailed paper on this SN, focusing on modeling its light curve to infer the properties of its progenitor, will be presented by Pichay et al. (in prep; \citealt{Pichay_2024_21ukt}). 
%In general, however, this evolution is typical for an IIn SN: At early times the Lorentzian is created through narrow Hydrogen emission broadened in the wings by electron scattering due to high optical depths \citep{Huang_2018}, and at later times as the optical depth decreases one begins to see emission both from the unshocked circumstellar material and from the dense shell formed between the shocks \citep{Smith_2016}. 
%Given this combination of features in the spectra, it must mean that the He absorption and H emission are coming from completely different regions of the SN, with the He absorption likely coming from the ejecta.   We thus suggest that 2021ukt must have had some relatively extensive CSM even at early times to form these narrow profiles.

%--------End SN 2021ukt-----------

\subsection{SN 2021njk}\label{subsec: 21njk}
SN 2021njk was discovered by ATLAS on UT 2021-05-25 06:48:57 (MJD 59354.38) with a non-detection two days prior \citep{Tonry2021_njk} and classified by the GSP \cite{2021TNSCR1876....1H}. Given that the TNS classification spectrum, taken on UT 2021-05-30 09:51:19 (t$_{V_\text{max}}$ = $-$11.3d), was obtained by LCO, this spectrum is also present in our study. SNID indicates the best match to SN 2005bf for this spectrum, in line with the TNS classification report by \cite{2021TNSCR1876....1H}. SN 2005bf is regarded as a peculiar type Ib because it shows a light curve with two peaks and He lines with increasing velocity over time \citep{2005ApJ...633L..97T}. However, for the purpose of spectroscopic classification, we list SN 2005bf as normal SN Ib. \cite{2021TNSCR1876....1H} also found a match with SN 1999ex using SNID and Superfit. Although \cite{2021TNSCR1876....1H} claimed SN 1999ex as peculiar type Ib, \cite{modjaz2014optical} argued that SN 1999ex is a normal type Ib and not an intermediate Ib/c. Additionally, we also found top matches with other normal SN Ib (iPTF13bvn, SN 2001ai, SN 2009jf) via SNID. We also ran the spectrum closest to the V-band maximum date, taken on UT 2021-06-13 10:05:47 (t$_{V_\text{max}}$ = 2.7d) through SNID. Although SNID did not find a favored type, the top 6 template matches are normal Ib. Thus, we classify SN 2021njk as a normal SN Ib.

\subsection{SN 2021hen}\label{subsec: 21hen}
SN 2021hen was discovered by the Automatic Learning for the Rapid Classification of Events (ALeRCE) broker team \citep{Alerce2021} on UT 2021-03-28 08:07:19.004 (MJD 59301.54) in ZTF data with a non-detection four days prior \citep{Forster2021_hen}. SN 2021hen was initially classified as an SN I using a GSP-obtained spectrum by \cite{Pellegrino2021_SN2021hen_Classification1}, who found SNID matches to both type Ia and type Ib SNe. This SN was reclassified as an SN Ib by \cite{2022TNSCR1989....1S} using a spectrum obtained 2.5 days after the initial classification spectrum. We ran the spectrum closest to the V-band maximum light, taken on UT 2021-04-06 11:54:36 (t$_{V_\text{max}}$ = $-$1.2d) through SNID. SNID classified this spectrum as a normal SN Ib, with 37.2$\%$ of the templates with rlap $\ge$ rlapmin corresponding to this type. For our earliest spectrum taken on UT 2021-03-30 13:01:37 (t$_{V_\text{max}}$ = $-$8.2d), SNID classified this spectrum as normal Ib, with 41.8$\%$ of the templates with rlap $\ge$ rlapmin corresponding to this type. \cite{Pellegrino2021_SN2021hen_Classification1} used the same spectrum for their initial classification on TNS, reporting SN 2021hen as type Ia or type Ib. But for our classification based on the same SN spectrum, SNID found the best match with SN Ib by using the updated MGS SNID template library with significantly more SESNe templates.

%--------End of SN 2021hen----------

\subsection{SN 2020hvp}\label{subsec: 20hvp}
SN 2020hvp was discovered by ATLAS on UT 2020-04-21 12:24:28 (MJD 58960.50) with a non-detection two days prior \citep{Tonry2020_hvp}. It was classified as an SN Ib by both \cite{2020TNSCR1129....1B} and \cite{2020TNSCR1523....1D} using spectra obtained 2 and 4 days after discovery, respectively. We ran our spectrum closest to the V-band maximum, taken on UT 2020-05-06 14:08:49 (t$_{V_\text{max}}$ = $-$0.4d) through SNID. SNID found that the best match is a normal SN Ib, with 29.3$\%$ of the templates with rlap $\ge$ rlapmin corresponding to this type. For the earliest spectrum taken on UT 2020-04-23 17:50:18 (t$_{V_\text{max}}$ = $-$13.2d), SNID did not find any favored type or subtype, but the top 6 template matches are normal SNe Ib. 

%--------End of SN 2020hvp----------

\subsection{SN 2019odp: Transitioning from SN Ic-bl to SN Ib}\label{subsec: 19odp}
SN 2019odp is an interesting transitional object, which shows features like SNe Ic-bl before the V-band maximum light (t$_{V_\text{max}} <$ 0) but transitions to type Ib afterwards (see also \citealt{Schweyer_2023_19odp} with their own set of spectra for this SN). SN 2019odp was discovered \citep{Nordin2019_odp} by ZTF on UT 2019-08-21 09:19:53 (MJD 58716.39) with a non-detection 7 days prior and was initially classified as SNe Ic-bl on UT 2019-08-23 05:23:25 (MJD = 58718.2) by \cite{2019TNSCR1595....1B} as part of the ePESSTO+ survey \citep{2015A&A...579A..40S}. Using presumably the same spectrum, \cite{Schweyer_2023_19odp} found that the spectrum was observed 16 days before the g-band peak (t$_{g_\text{max}}$ = $-$16d) and that it matches SNe Ic-bl. However, they reclassified SN 2019odp as type Ib based on the spectrum taken 26 days after the g-band peak (t$_{g_\text{max}}$ = 26d). Thus, \cite{Schweyer_2024_19odpClassification} updated the classification of SN 2019odp as type Ib on TNS. We ran the early spectra and the spectrum closest to the V-band peak date through SNID. SNID found SN 2007bi, a hydrogen-poor super-luminous SN Ic (SLSNe Ic; \citealt{Liu17}), as the best match for the earliest spectrum (t$_{V_\text{max}}$ = $-$15.7d). For the subsequent three early spectra (t$_{V_\text{max}}$ = $-$11.6d, $-$8.7d, $-$5.6d) , SNID found the best match with SN 1999bw (SNe Ic-bl; \citealt{2001ApJ...555..900P}). \cite{Liu17} found that the average absorption velocities are similar for SLSNe Ic and SNe Ic-bl, suggesting high velocity features in the earliest spectrum of SN 2019odp. The best match for our spectrum at $-$2.7d is a normal SN Ic, SN 2013ge.  For the spectrum closest to the V-band maximum light, taken on UT 2019-09-10 13:47:02 (t$_{V_\text{max}}$ = 1.4d), SNID found that the best-fitting type is normal Ib, with 55.9$\%$ of the templates with rlap $\ge$ rlapmin corresponding to this type. Thus, we reclassify SN 2019odp as a normal SN Ib based on spectra very close to maximum light (at t$_{V_\text{max}}$ = 1.4d) -- three weeks earlier than \cite{Schweyer_2023_19odp} did --  and indicate its transitional nature in Tables \ref{tab:sn_summary} and \ref{tab:tns} with "Ic-bl -> Ib". This transition from a SN Ic-bl to a SN Ib has been observed once before, in the famous SN 2008D, whose spectra resembled that of type Ic-bl shortly after explosion, that is, 15–10 days before the date of maximum light, but developed narrow absorption lines of helium by the date of maximum light (\citealt{Soderberg08,2008Sci...321.1185M, Modjaz_2009_08D}), similar to SN 2019odp.

%--------End of SN 2019odp----------

\subsection{SN 2016bau} \label{subsec: 16bau}
SN 2016bau was discovered on UT 2016-03-13 23:22:33 (MJD 57460.97) with a non-detection 9 days prior \citep{Arbour2016}. It was classified as an SN Ib on TNS using a classification spectrum obtained one day after discovery \citep{2016TNSCR.224....1B}. We ran our spectrum closest to the V-band maximum, taken on UT 2016-03-29 10:25:44 (t$_{V_\text{max}}$ = $-$1.7d) through SNID. SNID found that the best-fitting type is normal SN Ib, with 48.1$\%$ of the templates with rlap $\ge$ rlapmin corresponding to this type. For the earliest spectrum taken on UT 2016-03-15 05:24:59 (t$_{V_\text{max}}$ = $-$15.9d), SNID found that the best-fitting type is normal SN Ib, with 62.5\% of the templates with rlap $\ge$ rlapmin corresponding to this type. We note that \cite{Aryan_2021_16bau} published their own extensive dataset on this SN (including eight optical spectra, two of which were taken before maximum light), with a date of maximum calculation that is consistent with ours (see Section \ref{sec:peak_calc}). 

%--------End of SN 2016bau----------
%--------End of SN classifications----------

\section{Constructing SNID templates from our sample of young SNe Ib, including published spectra} \label{sec:snid_templates}
In order to improve the classification by SNID, it is crucial to enrich the SNID library by adding more classified SN templates of young SNe Ib. We created SNID templates using the 82 photospheric spectra of the eight SNe Ib published in this work. In addition, we included the 19 spectra of SN 2019odp published in \cite{Schweyer_2023_19odp} and 6 spectra of SN 2016bau published in \cite{Aryan_2021_16bau} when creating the SNID templates of those SNe. We note that all spectra in our sample are labeled as type Ib in our SNID templates, with the exception of type-changing SNe, such as SN 2022nyo, SN 2021ukt, and SN 2019odp, which are labeled as Ib-pec (peculiar type Ib). These objects are classified as peculiar type Ib due to their type-changing behavior. We also emphasize that the label "Ib-pec" in the SNID library can refer to various peculiarities (peculiar light curve behavior, changing type over time, etc.) shown by a SN. We encourage readers to explore the details of these specific SNe to better understand the unique characteristics that lead to their classification as peculiar. 

\par We used the logwave\footnote{\url{https://people.lam.fr/blondin.stephane/software/snid/howto.html\#logwave}} function in SNID to create spectral templates that are compatible with SNID 5.0. Our work adds 38 new SNe Ib spectra with $t_{V_{\text{max}}}$ < 0 to the SNID library, increasing the early SNe Ib spectra by $\sim$43$\%$. Including previously published spectra for SN 2019odp and SN 2016bau as mentioned above, the total increase is $\sim$54$\%$. SNID templates for the SNe Ib presented in this paper are publicly available via the METAL GitHub repository\footnote{\url{https://github.com/metal-sn/SESNtemple/tree/master}}, along with the previously published Modjaz Group sample (MGS) SNID templates, including those of young SNe Ic from \citet{Williamson2023YoungSNIc}. % will add link to the repo

%\par \textbf{We note that all spectra in our sample are labeled as type Ib in our SNID templates, regardless of any type-changing SNe present, as the primary objective of our dataset is to demonstrate the diversity observed in young SNe Ib. By consistently labeling these spectra as Ib, we aim to enrich the SNID library with early-time spectra of SNe Ib, providing a valuable reference for comparing newly observed young SNe with our sample. This consistent labeling also facilitates a better understanding of the evolution and characteristics of SNe Ib during their early stages post-explosion.}

%\par \textbf{We also acknowledge the limitations of the current SNID library, particularly its incompleteness regarding early spectra, which has posed challenges in finding suitable matches for several early spectra in our sample, as discussed in Section \ref{sec:classification}. To enhance the reliability of our classifications, we employ complementary methods, including the visual confirmation of key spectral features and the fitting of line profiles, as demonstrated in the case of SN 2021ukt (Section \ref{subsec: 21ukt}). We hope that our newly added SNID templates will help address some of the incompleteness in the library for early spectra of SNe Ib.}

%------- End of SNID templates-----------------------------

\section{Conclusions}
\par In the era of transient surveys with extensive sky coverage and rapid cadences such as ATLAS and ZTF, and with LSST on the horizon, the prompt and accurate classification of young SNe holds significant importance for subsequent follow-up observations. In this paper, we present a dataset of 82 photospheric spectra of eight SNe Ib, which includes a total of 38 spectra observed before the V-band maximum. This spectral dataset is publicly available via WISeREP. This dataset increases the number of published SNe Ib with at least three pre-maximum spectra by $\sim$60$\%$ and the number of available pre-maximum SNe Ib spectra in the SNID template library by $\sim$43$\%$. 

Our findings have a number of implications for the identification efforts of the SN community: \\

1) As mentioned above, almost half of our sample had a different type than listed in TNS or an evolving type over time, similar to the sample of young SNe Ic we published in \citet{Williamson2023YoungSNIc}. Thus, we suggest that the community keep observing and classifying SNe Ib over time when they are discovered young - and in general young CCSNe, including young SNe Ic -  so that the community can be alerted about any SNe that change types or develop peculiarities. In addition, since three of the eight SNe our sample had time-evolving types (SN 2019odp, SN 2021ukt, and 2022nyo), with additional ones in the literature (e.g., SN 2008D and SN 2022crv) we suggest that alternative SN identification methods be used that can capture the SN's amount of transitioning between different SESN subclasses - one such method was suggested by us in \citet{Williamson2019}, where we developed a PCA method for the photospheric spectra of SESNe that can naturally capture transition SNe (as applied to SN~2022crv, see Figure 17 in \citealt{Dong_2024_22crv}).

2) Incorporating early SN spectra into classification libraries as well as training sets for machine learning tools can improve the classification of young SNe. We suggest that the spectra of young SNe Ib, such as those presented here, be included in the spectral libraries of the various SN identification codes, such as SNID, Superfit, and GELATO. For the benefit of the community, we produce and make public the SNID templates of our young SNe Ib spectral dataset in this paper and release them via our METAL GitHub repository, formerly known as the SNYU github repository. 

Our release of these young spectra, which offer information about the outermost layers of SNe Ib ejecta, will hopefully facilitate work on constraining the characteristics and mass-loss mechanism of the SN progenitor. Additionally, these spectra will enable investigating different spectral characteristics, such as measuring line velocities and strength, to quantify diversity in SNe Ib at early times.

\begin{acknowledgements}
      \par This work makes use of observations from the Las Cumbres Observatory global telescope network.  M.M. and the METAL group at UVa acknowledge support in part from ADAP program grant No. 80NSSC22K0486, from the NSF grant AST-2206657 and from the HST GO program HST-GO-16656.
     The Las Cumbres Observatory group is supported by NSF grants AST-1911225 and AST-1911151. The SALT spectrum of SN 2022nyo was obtained through Rutgers University program 2022-1-MLT-004 (PI: SWJ). \\
      \textit{Software:} \texttt{Astropy} \citep{2018AJ....156..123A}, Matplotlib \citep{4160265}, Numpy \citep{2020Natur.585..357H}. Additionally, TeXGPT, a variant of GPT language model \citep{noauthororeditor} for LaTeX assistance on Overleaf, is used as an editing tool. 
\end{acknowledgements}

% WARNING
%-------------------------------------------------------------------
% Please note that we have included the references to the file aa.dem in
% order to compile it, but we ask you to:
%
% - use BibTeX with the regular commands:
%   \bibliographystyle{aa} % style aa.bst
%   \bibliography{Yourfile} % your references Yourfile.bib
%
% - join the .bib files when you upload your source files
%-------------------------------------------------------------------
% Place bibliography command where you want the bibliography to appear
\bibliographystyle{aa} % Set bibliography style
\bibliography{sources} % Specify your .bib file name without extension

\begin{appendix}
\onecolumn
\section{Spectral time-series plots}\label{sec:full_spectra}
%-----------Begin Time-series Plots----------
\begin{figure*}[!htb]
    \centering
    \begin{minipage}[b]{0.40\linewidth}
        \centering
        \includegraphics[width=\linewidth]{figures/SN2023ljf_crop_updated.png}
    \end{minipage}
    \hspace{0.05\linewidth} % Adjust the horizontal space as needed
    \begin{minipage}[b]{0.40\linewidth}
        \centering
        \includegraphics[width=\linewidth]{figures/SN2022nyo_crop_updated.png}
    \end{minipage}
    \vspace{0.05\linewidth} % Adjust the vertical space as needed
    \begin{minipage}[b]{0.40\linewidth}
        \centering
        \includegraphics[width=\linewidth]{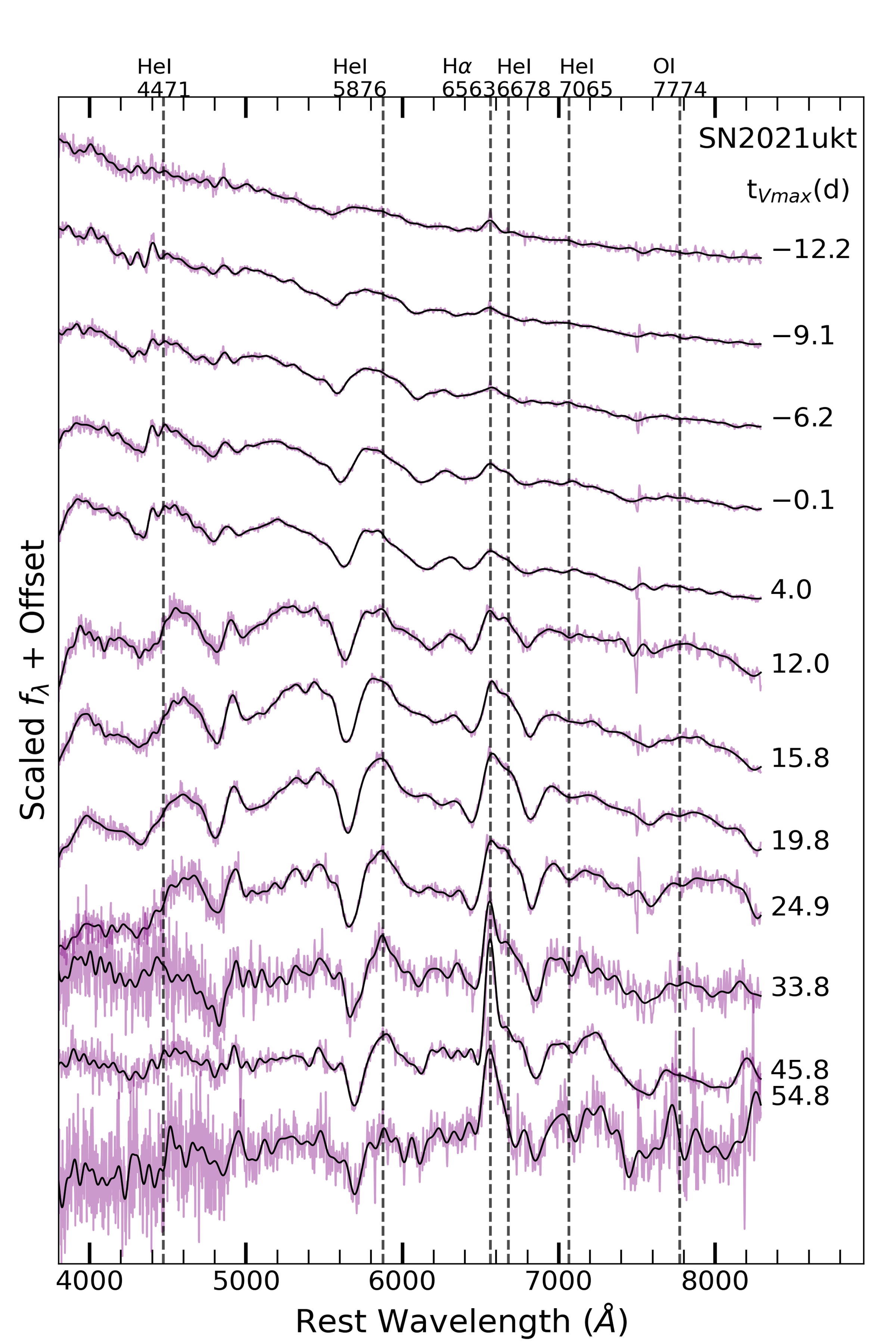}
    \end{minipage}
    \hspace{0.05\linewidth} % Adjust the horizontal space as needed
    \begin{minipage}[b]{0.40\linewidth}
        \centering
        \includegraphics[width=\linewidth]{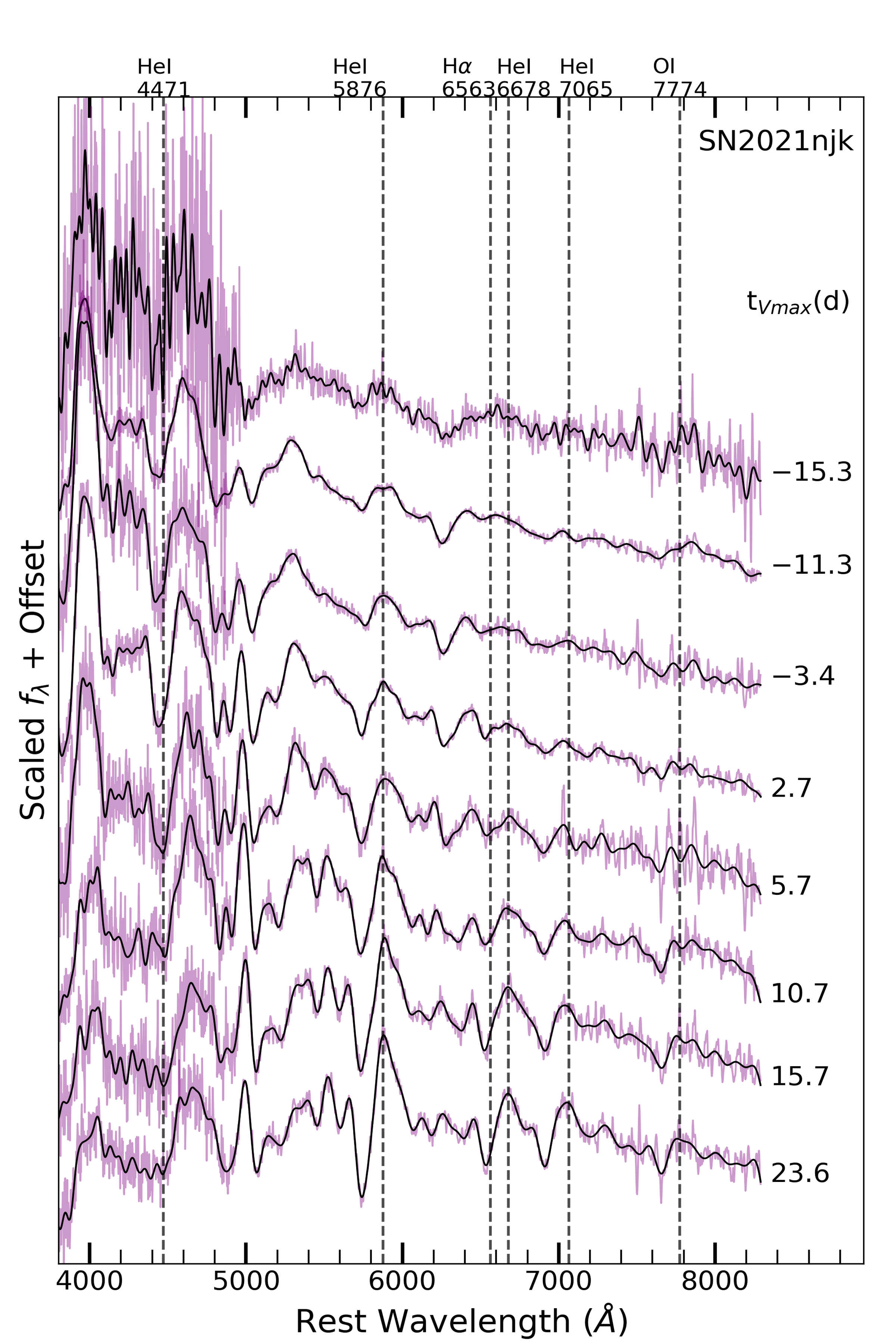}
    \end{minipage}
\end{figure*}
%\newpage
\begin{figure*}[!htb]
    \centering
    \begin{minipage}[b]{0.38\linewidth}
        \centering
        \includegraphics[width=\linewidth]{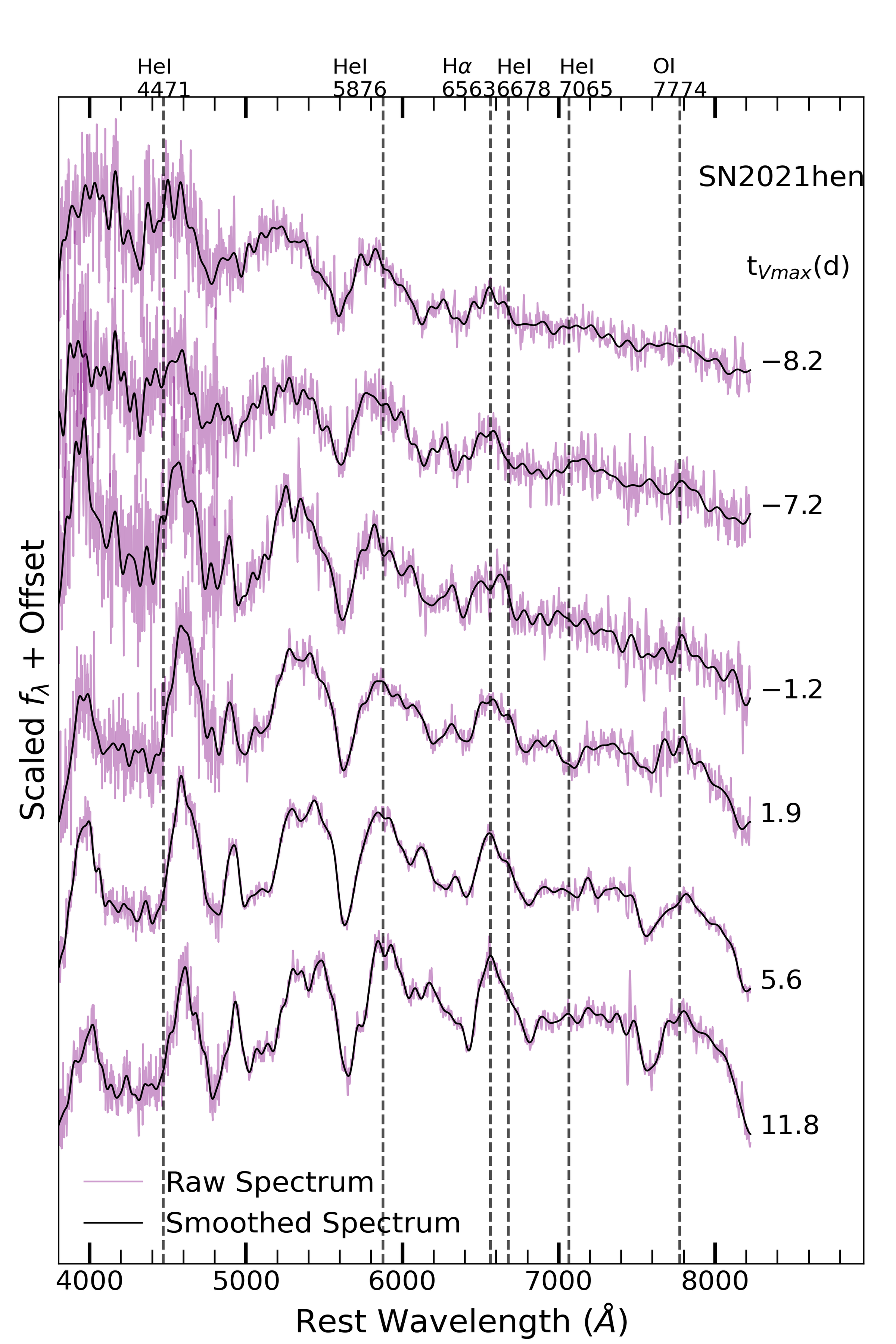}
    \end{minipage}
    \hspace{0.05\linewidth} % Adjust the horizontal space as needed
    \begin{minipage}[b]{0.38\linewidth}
        \centering
        \includegraphics[width=\linewidth]{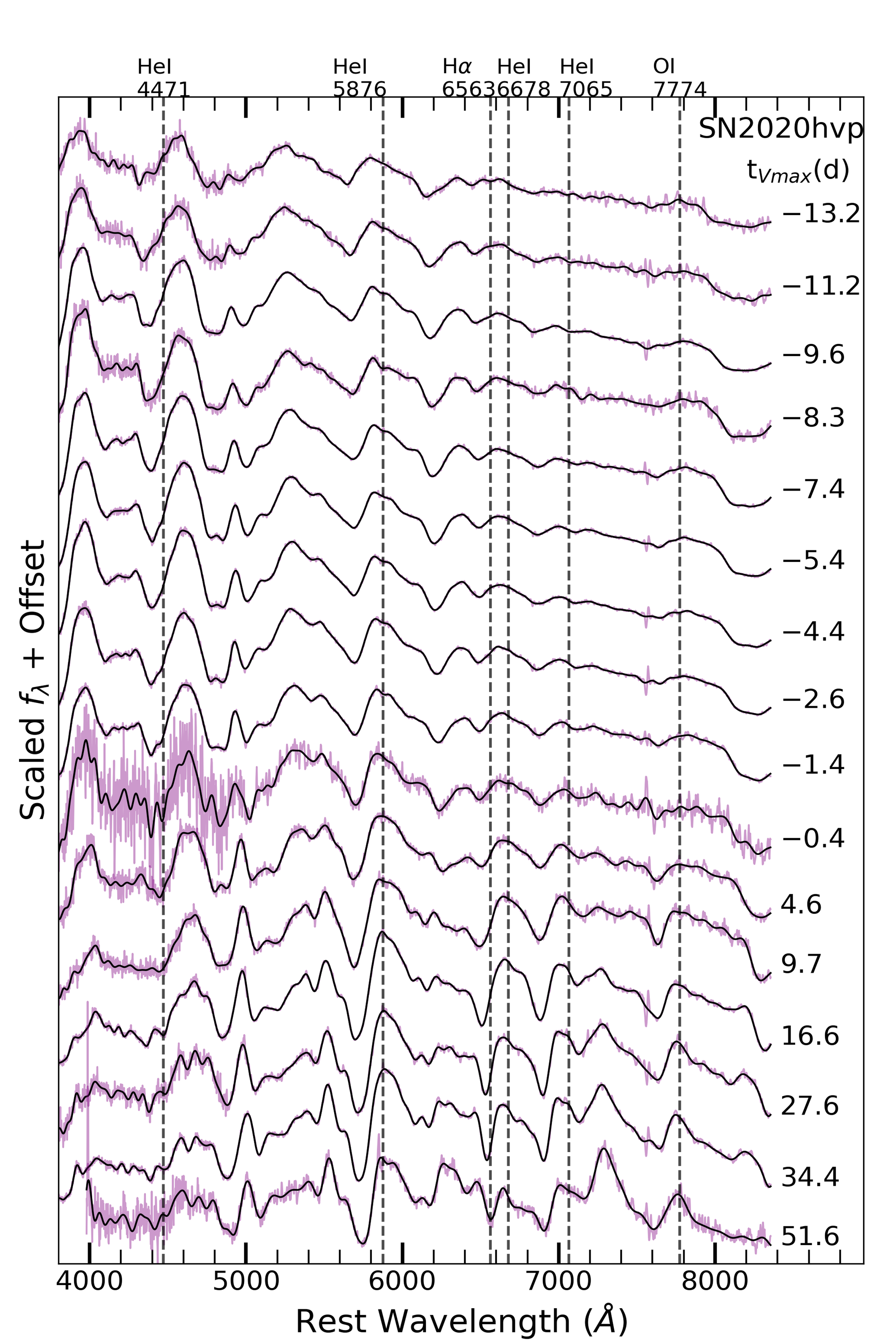}
    \end{minipage}
    
    \begin{minipage}[b]{0.38\linewidth}
        \centering
        \includegraphics[width=\linewidth]{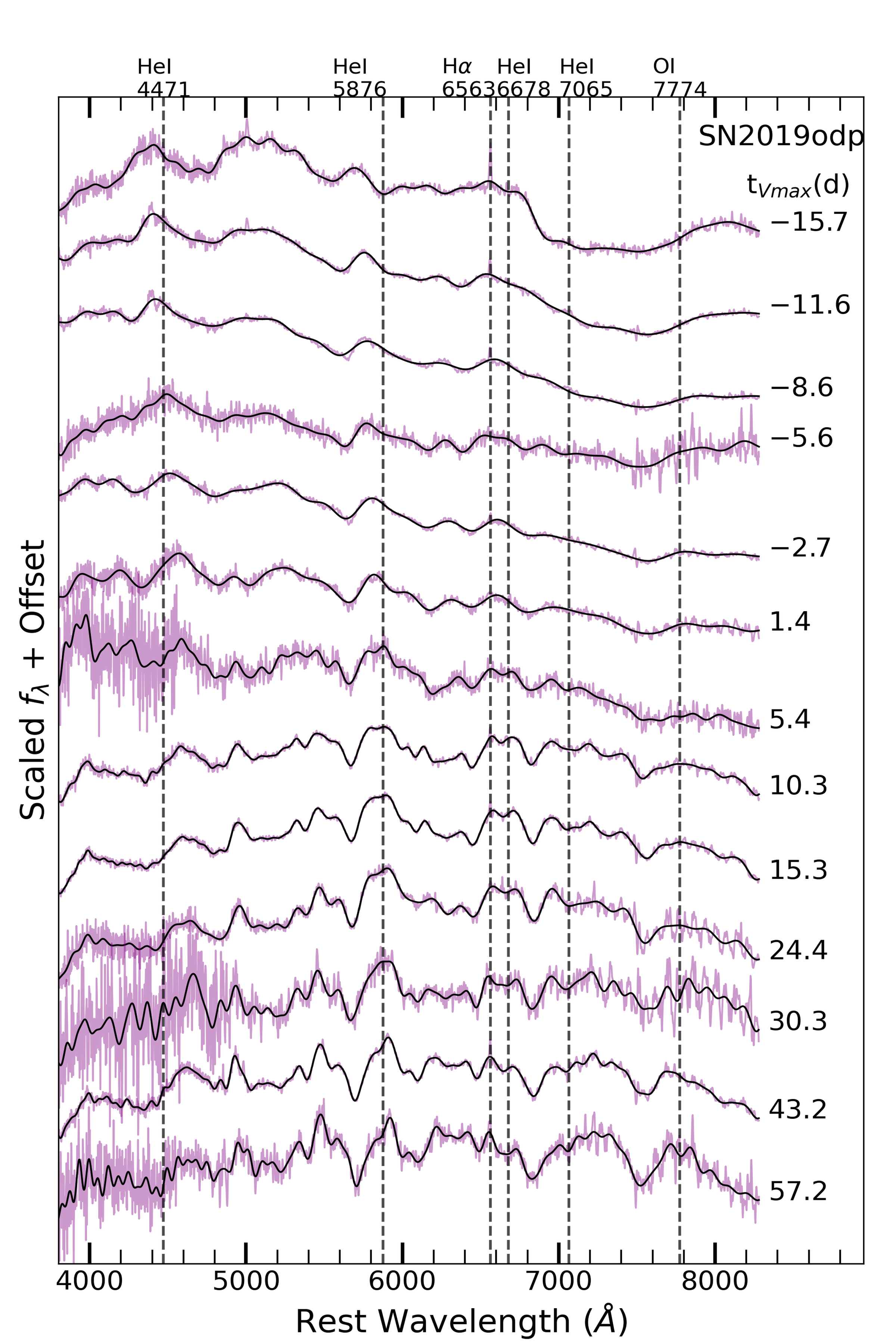}
    \end{minipage}
    \hspace{0.05\linewidth} % Adjust the horizontal space as needed
    \begin{minipage}[b]{0.38\linewidth}
        \centering
        \includegraphics[width=\linewidth]{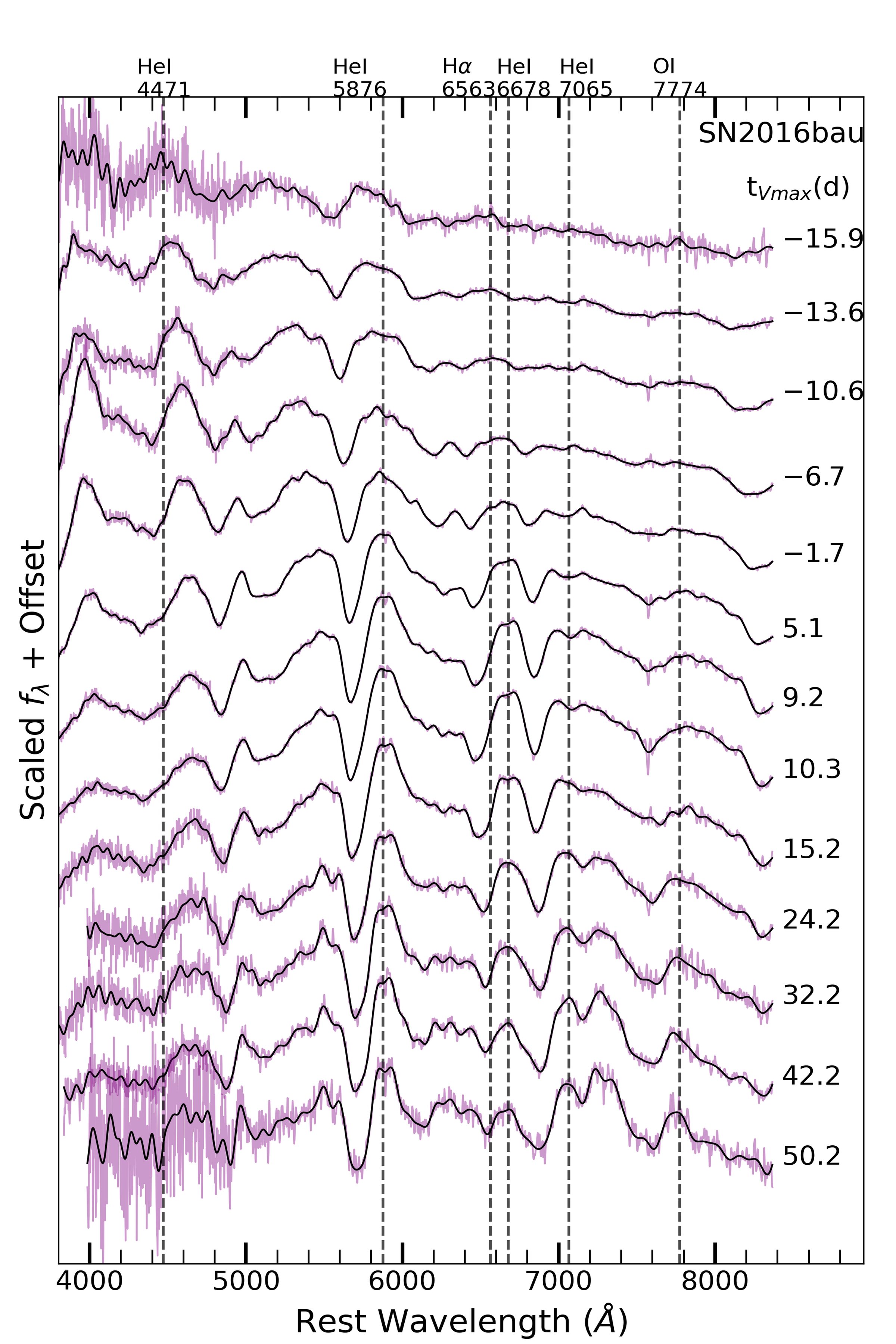}
    \end{minipage}
    \caption{Spectral time series for all eight SNe Ib presented in this work. The flux \( f_{\lambda} \), measured in units of \(\text{erg} \, \text{s}^{-1} \, \text{cm}^{-2} \, \text{Å}^{-1}\), is normalized, and an offset is added to enhance the visualization of the temporal evolution of the spectra. The smoothed spectra (produced following the procedure described in Appendix B of \citealt{Liu16}), shown as solid black lines, are highlighted in the foreground, while the reduced spectra before smoothing are in purple in the background. Dashed vertical lines mark the HeI 4471, HeI 5876, HeI 6678, HeI 7065, H$\alpha$ 6563, and OI 7774 lines, and the phases are relative to the date of the V-band maximum ($t_{V_\text{max}}$). We note that we manually removed galaxy emission lines from the spectra at $t_{V_\text{max}}$ = $-$13.8d and $-$4.1d of SN 2022nyo. All SNe were chosen to have at least three spectra before maximum light in V-band. Thus, the observed change in the SN types for almost half of our sample could be uncovered with our large number of pre-maximum and post-maximum spectra. See Figure~\ref{fig:illustrative_spectra} in the main text for an illustrative example.}
    \label{fig:full_spectra}
\end{figure*}
% ------------------END Time-series Figure------------------

\onecolumn
\section{Observation log}\label{sec: obs}
\begin{longtable}{p{3.0cm} p{0.8cm} p{3.5cm} p{3.8cm} c p{2.0cm} p{1.0cm} p{1.0cm}}
\caption{Spectra observation log of young type Ib SNe.}\label{tab: obs_log}\\
\hline
\textbf{SN (t$_{V_{\text{max}}}$ days)} & \textbf{type} & \textbf{Date of Observation} & \textbf{Telescope}/\textbf{Instrument} & \textbf{Exposure (s)} & \textbf{Slit ($\arcsec$)} & \textbf{Airmass} \\
\hline
\endfirsthead
\caption{continued.}\\
\hline
\textbf{SN (t$_{V_{\text{max}}}$ days)} & \textbf{type} & \textbf{Date of observation} & \textbf{Telescope}/\textbf{Instrument} & \textbf{Exposure (s)} & \textbf{Slit ($\arcsec$)} & \textbf{Airmass} \\
\hline
\endhead
\hline
\endfoot
\hline
\endlastfoot
SN 2023ljf($-$14.5) & Ib & 2023-06-23 09:29:05 & OGG 2m/FLOYDS & 3600 & 2\arcsec & 1.42 \\
SN 2023ljf($-$12.5) & Ib & 2023-06-25 08:16:46 & OGG 2m/FLOYDS & 3600 & 2\arcsec & 1.42 \\
SN 2023ljf($-$10.5) & Ib & 2023-06-27 08:21:33 & OGG 2m/FLOYDS & 2700 & 2\arcsec & 1.19 \\
SN 2023ljf($-$7.6) & Ib & 2023-06-30 06:25:18 & OGG 2m/FLOYDS & 2700 & 2\arcsec & 1.04 \\
SN 2023ljf($-$4.6) & Ib & 2023-07-03 07:23:36 & OGG 2m/FLOYDS & 2700 & 2\arcsec & 1.12 \\
SN 2023ljf(19.4) & Ib & 2023-07-27 07:20:07 & OGG 2m/FLOYDS & 2700 & 2\arcsec & 1.4 \\
SN 2023ljf(25.4) & Ib & 2023-08-02 05:57:19 & OGG 2m/FLOYDS & 2700 & 2\arcsec & 1.12 \\
SN 2023ljf(28.4) & Ib & 2023-08-05 06:20:32 & OGG 2m/FLOYDS & 2700 & 2\arcsec & 1.29 \\
SN 2023ljf(40.4) & Ib & 2023-08-17 05:59:48 & OGG 2m/FLOYDS & 2700 & 2\arcsec & 1.41 \\
\midrule
SN 2022nyo($-$13.8)\tablenotemark{a} & IIb & 2022-07-03 01:55:28 & SOAR~4.1m/GHTS~BLUE & 1800 & 1\arcsec & 1.43 \\
SN 2022nyo($-$4.1)\tablenotemark{a} & Ib & 2022-07-12 18:29:14 & SALT/RSS & 1533 & 1.5\arcsec & 1.27\\
SN 2022nyo($-$1.5) & Ib & 2022-07-15 10:06:09 & COJ 2m/FLOYDS & 2700 & 2\arcsec & 1.4\\
SN 2022nyo(23.4) & Ib & 2022-08-09 08:30:45 & COJ 2m/FLOYDS & 2700 & 2\arcsec & 1.42\\
SN 2022nyo(33.4) & Ib & 2022-08-19 08:36:22 & COJ 2m/FLOYDS & 2700 & 2\arcsec & 1.68\\
\midrule
SN 2021ukt($-$12.2) & IIn & 2021-08-03 11:41:04 & OGG 2m/FLOYDS & 1200 & 2\arcsec & 1.39\\
SN 2021ukt($-$9.1) & Ib & 2021-08-06 12:17:32 & OGG 2m/FLOYDS & 1200 & 2\arcsec & 1.21\\
SN 2021ukt($-$6.2) & Ib & 2021-08-09~11:57:22 & OGG 2m/FLOYDS & 1200 & 2\arcsec & 1.24\\
SN 2021ukt($-$0.1) & Ib & 2021-08-15 13:30:07 & OGG 2m/FLOYDS & 1200 & 2\arcsec & 1.08\\
SN 2021ukt(4.0) & Ib & 2021-08-19 14:38:17 & OGG 2m/FLOYDS & 1200 & 2\arcsec & 1.16\\
SN 2021ukt(12.0) & Ib & 2021-08-27 15:10:45 & COJ 2m/FLOYDS & 1200 & 2\arcsec & 1.2\\
SN 2021ukt(15.8) & Ib & 2021-08-31 10:25:41 & OGG 2m/FLOYDS & 1199 & 2\arcsec & 1.25\\
SN 2021ukt(19.8) & Ib & 2021-09-04 12:08:46 & OGG 2m/FLOYDS & 1200 & 2\arcsec & 1.08\\
SN 2021ukt(24.9) & Ib & 2021-09-09 12:43:45 & COJ 2m/FLOYDS & 2700 & 2\arcsec & 1.47\\
SN 2021ukt(33.8) & Ib & 2021-09-18 10:51:57 & OGG 2m/FLOYDS & 3600 & 2\arcsec & 1.09\\
SN 2021ukt(45.8) & Ib & 2021-09-30 09:59:09  & OGG 2m/FLOYDS & 3600 & 2\arcsec & 1.09\\
SN 2021ukt(54.8) & Ib & 2021-10-09 10:46:54 & COJ 2m/FLOYDS & 3600 & 2\arcsec & 1.44\\
\midrule
SN 2021njk($-$15.3) & Ib & 2021-05-26 10:39:13 & COJ 2m/FLOYDS & 2700 & 2\arcsec & 1.20\\
SN 2021njk($-$11.3) & Ib & 2021-05-30 09:51:19 & COJ 2m/FLOYDS & 3600 & 2\arcsec & 1.14\\
SN 2021njk ($-$3.4) & Ib & 2021-06-07 08:07:57 & COJ 2m/FLOYDS & 1800 & 2\arcsec & 1.02\\
SN 2021njk(2.7) & Ib & 2021-06-13 10:05:47 & COJ 2m/FLOYDS & 1800 & 2\arcsec & 1.3\\
SN 2021njk(5.7) & Ib & 2021-06-16 10:20:45 & COJ 2m/FLOYDS & 1800 & 2\arcsec & 1.42\\
SN 2021njk(10.7) & Ib & 2021-06-21 09:43:39 & COJ 2m/FLOYDS & 1800 & 2\arcsec & 1.34\\
SN 2021njk(15.7) & Ib & 2021-06-26 09:47:11 & COJ 2m/FLOYDS & 2700 & 2\arcsec & 1.51\\
SN 2021njk(23.6) & Ib & 2021-07-04 09:31:19 & COJ 2m/FLOYDS & 2700 & 2\arcsec & 1.61\\
\midrule
SN 2021hen($-$8.2) & Ib & 2021-03-30 13:01:37 & OGG 2m/FLOYDS & 3600 & 2\arcsec & 1.24\\
SN 2021hen($-$7.2) & Ib & 2021-03-31 11:53:45 & OGG 2m/FLOYDS & 3600 & 2\arcsec & 1.08\\
SN 2021hen($-$1.2) & Ib & 2021-04-06 11:54:36 &  OGG 2m/FLOYDS & 3600 & 2\arcsec & 1.12\\
SN 2021hen(1.9) & Ib & 2021-04-09 14:13:33 & COJ 2m/FLOYDS & 3600 & 2\arcsec & 1.85\\
SN 2021hen(5.6) & Ib & 2021-04-13 08:26:33 & OGG 2m/FLOYDS & 3600 & 2\arcsec & 1.05\\
SN 2021hen(11.8) & Ib & 2021-04-19 12:16:47 & OGG 2m/FLOYDS & 3600 & 2\arcsec & 1.38\\
\midrule
SN 2020hvp($-$13.2) & Ib & 2020-04-23 17:50:18 & COJ 2m/FLOYDS & 3600 & 2\arcsec & 1.34\\
SN 2020hvp($-$11.2) & Ib & 2020-04-25 17:41:11 & COJ 2m/FLOYDS & 1800 & 2\arcsec & 1.28\\
SN 2020hvp($-$9.6) & Ib & 2020-04-27 09:20:25  & OGG 2m/FLOYDS & 1800 & 2\arcsec & 1.48\\
SN 2020hvp($-$8.3) & Ib & 2020-04-28 16:41:04 & COJ 2m/FLOYDS & 1200 & 2\arcsec & 1.17\\
SN 2020hvp($-$7.4) & Ib & 2020-04-29 13:46:48 & OGG 2m/FLOYDS & 1200 & 2\arcsec & 1.20\\
SN 2020hvp($-$5.4) & Ib & 2020-05-01 12:22:17 & OGG 2m/FLOYDS & 1200 & 2\arcsec & 1.09\\
SN 2020hvp($-$4.4) & Ib & 2020-05-02 14:08:36 & OGG 2m/FLOYDS & 900 & 2\arcsec & 1.30\\
SN 2020hvp($-$2.6) & Ib & 2020-05-04 08:53:23 & OGG 2m/FLOYDS & 900 & 2\arcsec & 1.52\\
SN 2020hvp($-$1.4) & Ib & 2020-05-05 12:57:46 & OGG 2m/FLOYDS & 900 & 2\arcsec & 1.15\\
SN 2020hvp($-$0.4) & Ib & 2020-05-06 14:08:49 & COJ 2m/FLOYDS & 900 & 2\arcsec & 1.20\\
SN 2020hvp(4.6) & Ib & 2020-05-11 13:07:53 & COJ 2m/FLOYDS & 900 & 2\arcsec & 1.29\\
SN 2020hvp(9.7) & Ib & 2020-05-16 16:25:52 & COJ 2m/FLOYDS & 900 & 2\arcsec & 1.27\\
SN 2020hvp(16.6) & Ib & 2020-05-23 12:49:18 & OGG 2m/FLOYDS & 1800 & 2\arcsec & 1.35\\
SN 2020hvp(27.6) & Ib & 2020-06-03 12:15:10 & OGG 2m/FLOYDS & 1800 & 2\arcsec & 1.39\\
SN 2020hvp(34.4) & Ib & 2020-06-10 07:46:35 & OGG 2m/FLOYDS & 2400 & 2\arcsec & 1.17\\
SN 2020hvp(51.6) & Ib & 2020-06-27 13:34:03 & COJ 2m/FLOYDS & 2700 & 2\arcsec & 1.30\\
\midrule
SN 2019odp($-$15.7) & Ic-bl & 2019-08-24 12:44:36 & OGG 2m/FLOYDS & 3600 & 2\arcsec & 1.14\\
SN 2019odp($-$11.6) & Ic-bl & 2019-08-28 13:33:07 & OGG 2m/FLOYDS & 3599 & 2\arcsec & 1.38\\
SN 2019odp($-$8.7) & Ic-bl & 2019-08-31 13:35:03 & OGG 2m/FLOYDS & 3600 & 2\arcsec & 1.46\\
SN 2019odp($-$5.6) & Ic-bl & 2019-09-03 13:46:45 & COJ 2m/FLOYDS & 3600 & 2\arcsec & 1.43\\
SN 2019odp($-$2.7) & Ic-bl & 2019-09-06 11:08:12 & OGG 2m/FLOYDS & 2700 & 2\arcsec & 1.04\\
SN 2019odp(1.4) & Ib & 2019-09-10 13:47:02 & COJ 2m/FLOYDS & 2700 & 2\arcsec & 1.43\\
SN 2019odp(5.4) & Ib & 2019-09-14 14:48:18 & COJ 2m/FLOYDS & 2700 & 2\arcsec & 1.58\\
SN 2019odp(10.3) & Ib & 2019-09-19 12:10:09 &  OGG 2m/FLOYDS & 2700 & 2\arcsec & 1.35\\
SN 2019odp(15.3) & Ib & 2019-09-24 11:23:11 & OGG 2m/FLOYDS & 2700 & 2\arcsec & 1.23\\
SN 2019odp(24.4) & Ib & 2019-10-03 13:29:53 & COJ 2m/FLOYDS & 2700 & 2\arcsec & 1.57\\
SN 2019odp(30.3) & Ib & 2019-10-09 12:35:08 & COJ 2m/FLOYDS & 2700 & 2\arcsec & 1.48\\
SN 2019odp(43.2) & Ib & 2019-10-22 09:12:29 & OGG 2m/FLOYDS & 3600 & 2\arcsec & 1.2\\
SN 2019odp(57.2) & Ib & 2019-11-05 10:23:11 & COJ 2m/FLOYDS & 3600 & 2\arcsec & 1.46\\
\midrule
SN 2016bau($-$15.9) & Ib & 2016-03-15 05:24:59 & OGG 2m/FLOYDS & 2700 & 2\arcsec & 1.99\\
SN 2016bau($-$13.6) & Ib & 2016-03-17 13:01:07 & OGG 2m/FLOYDS & 2700 & 2\arcsec & 1.56\\
SN 2016bau($-$10.6) & Ib & 2016-03-20 12:47:21 & OGG 2m/FLOYDS & 2700 & 2\arcsec & 1.55\\
SN 2016bau($-$6.7) & Ib & 2016-03-24 12:04:09 & OGG 2m/FLOYDS & 2700 & 2\arcsec & 1.44\\
SN 2016bau($-$1.7) & Ib & 2016-03-29 10:25:44 & OGG 2m/FLOYDS & 2700 & 2\arcsec & 1.25\\
SN 2016bau(5.1) & Ib & 2016-04-05 06:56:20 & OGG 2m/FLOYDS & 2700 & 2\arcsec & 1.25\\
SN 2016bau(9.2) & Ib & 2016-04-09 08:26:46 & OGG 2m/FLOYDS & 2700 & 2\arcsec & 1.19\\
SN 2016bau(10.3) & Ib & 2016-04-10 11:53:54 & OGG 2m/FLOYDS & 2700 & 2\arcsec & 1.7\\
SN 2016bau(15.2) & Ib & 2016-04-15 08:53:05 & OGG 2m/FLOYDS & 2700 & 2\arcsec & 1.22\\
SN 2016bau(24.2) & Ib & 2016-04-24 08:37:56 & OGG 2m/FLOYDS & 2700 & 2\arcsec & 1.24\\
SN 2016bau(32.2) & Ib & 2016-05-02 09:18:02 & OGG 2m/FLOYDS & 2700 & 2\arcsec & 1.4\\
SN 2016bau(42.2) & Ib & 2016-05-12 09:28:37 & OGG 2m/FLOYDS & 2700 & 2\arcsec & 1.59\\
SN 2016bau(50.2) & Ib & 2016-05-20 08:55:44 & OGG 2m/FLOYDS & 2700 & 2\arcsec & 1.59\\
\end{longtable}
\begin{minipage}{\textwidth}
\raggedright
%\centering
\tablefoot{
    \tablenotemark{a} Host galaxy HII region emission lines are removed from these spectra. Figure \ref{fig:illustrative_spectra} and \ref{fig:full_spectra} show the spectra without the galaxy emission lines.
}
%\textbf{Note:} Host galaxy HII region emission lines are removed from these spectra. Figure \ref{fig:spectra} shows the spectra without the galaxy emission lines.
\end{minipage}
%\begin{figure*}[b]
%\centering
%\caption*{%
%\textbf{Note:} Host galaxy HII region emission lines are removed from these spectra. Figure \ref{fig:spectra} shows the spectra without the galaxy emission lines.%
%}
\end{appendix}

\end{document}